\documentclass[letterpaper]{JHEP3}

\usepackage{amsmath}

\title{Stealths on Anisotropic Holographic Backgrounds}

\author{Eloy Ay\'on-Beato \\
Departamento de F\'isica, CINVESTAV-IPN, Apdo. Postal 14-740, 07000,
M\'exico D.F., M\'exico\\
Instituto de Ciencias F\'isicas y Matem\'aticas,
Universidad Austral de Chile, Casilla 567 Valdivia, Chile\\
Centro de Estudios Cient\'ificos (CECs),
Casilla 1468 Valdivia, Chile\\
E-mail: \email{ayon-beato-at-fis.cinvestav.mx}}

\author{Mokhtar Hassa\"{i}ne \\
Instituto de Matem\'aticas y F\'isica, Universidad de Talca, Casilla 747,
Talca, Chile\\
E-mail: \email{hassaine-at-inst-mat.utalca.cl}}

\author{Mar\'ia Montserrat Ju\'arez-Aubry \\
Departamento de F\'isica, CINVESTAV-IPN, Apdo. Postal 14-740, 07000,
M\'exico D.F., M\'exico\\
Instituto Tecnol\'ogico y de Estudios Superiores de Monterrey,
Campus Puebla, V\'ia Atlixc\'ayotl No. 2301,
Reserva Territorial Atlixc\'ayotl, Puebla. C.P. 72453 Puebla, M\'exico\\
E-mail: \email{mjuarez-at-fis.cinvestav.mx}}

\abstract{In this paper, we are interested in exploring the
existence of stealth configurations on anisotropic backgrounds
playing a prominent role in the non-relativistic version of the
gauge/gravity correspondence. By stealth configuration, we mean
a nontrivial scalar field nonminimally coupled to gravity whose
energy-momentum tensor evaluated on the anisotropic background
vanishes identically. In the case of a Lifshitz spacetime with
a nontrivial dynamical exponent $z$, we spotlight the role
played by the anisotropy to establish the holographic character
of the stealth configurations, i.e.\ the scalar field is shown
to only depend on the radial holographic direction. This
configuration which turns out to be massless and without
integration constants is possible for a unique value of the
nonminimal coupling parameter. Then, using a simple conformal
argument, we map this configuration into a stealth solution
defined on the so-called hyperscaling violation metric which is
conformally related to the Lifshitz spacetime. This holographic
configuration obtained through a conformal mapping constitutes
only a particular class within the stealth solutions defined on
the hyperscaling violation as it is shown by deriving the most
general stealth configurations. The case of the Schr\"odinger
background is also exhaustively analyzed and we establish that
the presence of the null direction makes their stealth
configurations not necessarily holographic in general and
characterized by a self-interacting behavior. Finally, for
completeness we also study the stealth configurations on the
Schr\"odinger inspired hyperscaling violation spacetimes.}

\begin{document}

\maketitle

\section{Introduction}

The fundamental tenet of General Relativity lies in the fact
that gravity is a manifestation of the curvature of spacetime
produced by the presence of matter sources. This phenomenon is
encoded in the equations proposed by Einstein a century ago
that relate a gravity tensor in the left hand side, which only
depends on the metric, to the energy-momentum tensor of the
matter source in the right hand side. Nevertheless, one can ask
the following question: can there exist background metrics and
nontrivial matter sources such that both sides of Einstein
equations vanish identically? This question can be reformulated
as follows: for a fixed gravitational background presenting
some physical interest, does a nontrivial matter source exist
such that its energy-momentum vanishes once evaluated on this
background? These kind of solutions of Einstein equations have
been dubbed stealth configurations and are characterized by
their lack of back-reaction on the gravitational field causing
an impossibility of curving the underlying geometry. Such
configurations have been derived previously using scalar fields
nonminimally coupled to gravity on the BTZ black hole
\cite{Banados:1992wn} in three dimensions
\cite{AyonBeato:2004ig}, and on higher-dimensional Minkowski
\cite{AyonBeato:2005tu} and (A)dS spacetimes
\cite{Ayon-Beato:SAdS}. Concretely for the conformal coupling,
they have been also exhibited on any homogeneous and isotropic
universe \cite{Ayon-Beato:2013bsa}.

In this work, we explore whether nonminimally coupled (and
possibly self-interacting) scalar fields may still be a good
laboratory to define stealth configurations on spacetimes
playing a prominent role in the gauge/gravity correspondence
\cite{Maldacena:1997re}. The paradigmatic example as
gravitational dual is the AdS space. The existence of the
stealth in this case has been established in
\cite{Ayon-Beato:SAdS}. Other interesting gravitational duals
are those occurring in the non-relativistic versions of the
gauge/gravity correspondence. Significant examples are those
where anisotropic scalings play an outstanding role such as the
Lifshitz spacetimes \cite{Kachru:2008yh} or their
generalizations exhibiting hyperscaling violation
\cite{Charmousis:2010zz}. Historically, the first example of
non-relativistic holography was provided with the Schr\"odinger
background, see \cite{Son:2008ye,Balasubramanian:2008dm}.
However, the presence of a (compact) null direction required to
ensure the Galilean invariance makes its holographic
interpretation a subject of debate.

The relevance of stealth configurations for gauge/gravity
duality lies in the fact that their fluctuations are not expected
to be stealth themselves.\footnote{We thank
C.~Terrero-Escalante for pointing out this non trivial issue.}
Correspondingly, the perturbations of the gravitational duals
could be modified by the presence of the stealth, and since
this weak field limit is in correspondence to the strongly
correlated regime of the dual quantum theory, the associated
holographic predictions could be potentially modified. In this
paper, we definitely establish the existence of stealth
configurations on all the anisotropic backgrounds studied in
the current literature, namely the Lifshitz and Schr\"odinger
backgrounds as well as for their hyperscaling violation
generalizations, and spotlight their similitude and their
differences. This paves the way to initiate an ensuing program
exploring the holographic consequences of the existence of the
stealths and their nontrivial perturbations in the
non-relativistic version of gauge/gravity duality.

The action we choose to model stealth configurations is the one
of a self-interacting scalar field $\Phi$ nonminimally coupled
to gravity and parameterized in terms of the nonminimal
coupling parameter $\xi$ as
\begin{equation}
S_{\xi}[\Phi,g_{\mu\nu}]=\int{d^{D}x}\,\sqrt{-g}\left(
-\frac{1}{2}\partial_{\mu}\Phi\,\partial^{\mu}\Phi
-\frac{1}{2}\xi{R}\,\Phi^2-U(\Phi)\right),
\label{eq:action}
\end{equation}
where $R$ stands for the scalar curvature and $U(\Phi)$
represents a possible self-interaction potential. As said
before, a stealth configuration is composed by two ingredients:
the given spacetime background (which, in our case, will be the
Lifshitz, the Schr\"odinger or their hyperscaling
generalizations) and the nontrivial field (given by the
nonminimally coupled scalar field) whose energy-momentum tensor
evaluated on the background vanishes. In this sense, the
related constraints can also be understood as if both
ingredients correspond to extrema of the matter action
describing just the field. In fact, fixing the background a
priori, the related conditions become only constraints for the
dynamical field. In our case, the conditions defining the
stealth, which also correspond to the variations of the action
(\ref{eq:action}), are given by
\begin{subequations}\label{eq:SEqs}
\begin{eqnarray}
T_{\alpha\beta}=\partial_{\alpha}\Phi\,\partial_{\beta}\Phi
-g_{\alpha\beta}\left(\frac{1}{2}\partial_{\mu}\Phi\,\partial^{\mu}\Phi
+U(\Phi)\right)
+\xi\left(g_{\alpha\beta}\Box-\nabla_{\alpha}\nabla_{\beta}
+G_{\alpha\beta}\right)\Phi^2&=&0,\qquad\label{eq:eqsTmunu}\\
\Box\Phi-\xi R\Phi-\frac{dU(\Phi)}{d\Phi}&=&0.\label{eq:eqsKG}
\end{eqnarray}
\end{subequations}
In fact, it is sufficient to solve only the first constraints
(\ref{eq:eqsTmunu}), since the fulfillment of the field
equation (\ref{eq:eqsKG}) for a nontrivial field is warranted
from the conservation of the energy-momentum tensor due to the
diffeomorphism invariance of action (\ref{eq:action}).

The paper is organized as follows. We start with a brief
introduction on non-relativistic holographic backgrounds in
Sec.~$2$. After that we prove in Sec.~$3$ that Lifshitz
spacetimes support stealth configurations by solving the
stealth constraints on the Lifshitz background in full
generality. We emphasize that a straightforward consequence of
its anisotropy is that, contrary to the AdS case
\cite{Ayon-Beato:SAdS}, the Lifshitz stealth is not only
stationary but must be also strictly holographic in nature in
the sense that it only admits a dependence on the holographic
direction. Other interesting features which differ radically
from the isotropic cases are that the nonminimal coupling
parameter takes a fixed value parameterized in terms of the
corresponding dynamical exponent for any dimension, and these
anisotropic configurations must be free of any
self-interaction. In Sec.~$4$, we exploit the conformal
relation between the Lifshitz and hyperscaling violation
metrics and we extend the stealth configuration to the last
spacetime using conformal arguments in arbitrary dimension.
Later on, we show that these configurations constitute only a
particular class of the possible stealths on the hyperscaling
violation metric by deriving its most general stealth
solutions. We exhibit special values of the exponents for which
the stealth depends on the transverse directions in addition to
the holographic coordinate. In Sec.~$5$ the Schr\"odinger case
is analyzed in detail. The presence of the null coordinate
makes the Schr\"odinger stealth configuration not necessarily
holographic. Finally, we also analyze the hyperscaling
violation case derived from the Schr\"odinger metric. The last
section is devoted to our conclusions.

\section{Non-relativistic holographic backgrounds}

In condensed matter physics, a quantum phase transition occurs
between two different phases at zero temperature. At this
critical point, the system becomes invariant under a scaling
symmetry with eventually different weights between space and
time,
\begin{equation}\label{eq:scaling}
t\mapsto\lambda^z\,t,\qquad \vec{x}\mapsto\lambda\vec{x},
\end{equation}
where the relative weight $z$ is called the dynamical critical
exponent. The quantum critical point can be very useful to have
a good comprehension of the entire phase diagram, and hence a
good analysis of this point can be of particular interest in
order to compute transport coefficients, like the electrical
conductivity in the case of superfluid-insulator phase
transition or some thermal quantities. In general, these
quantities are difficult to compute because the systems are
usually strongly coupled. The AdS/CFT correspondence, valid for
$z=1$, has been proved to be a very promising tool for studying
strongly coupled systems by mapping them into the weak regime
of classical theories of gravity and establishing a dictionary
between both systems \cite{Maldacena:1997re}. However, other
values of $z$ are present experimentally in particular for
different condensed matter phenomena. This is one of the main
motivations behind the recent interest of extending the AdS/CFT
correspondence to non-relativistic physics and to condensed
matter applications. In this perspective, there are two
symmetry groups playing an important role: the Schr\"odinger
group which may be viewed as the non-relativistic cousin of the
conformal group and the Lifshitz group which is characterized
by an anisotropic scaling symmetry but without the Galilean
boosts of the first example. For this latter case, the gravity
dual metric is referred to as the Lifshitz metric
\cite{Kachru:2008yh} and is given in $D$ dimensions by
\begin{equation}\label{eq:Lifshitzmetric}
ds_{\mathrm{L}}^2=-\frac{r^{2z}}{l^{2z}}dt^2+\frac{l^2}{r^2}dr^2
+\frac{r^2}{l^2}d\vec{x}^2,
\end{equation}
where $\vec{x}=(x^1,\ldots,x^{D-2})$. Indeed, it is simple to
check that the dynamical scalings (\ref{eq:scaling})
supplemented with an additional scaling in the holographic
direction, $r\mapsto\lambda^{-1}r$, act as an isometry for the
metric (\ref{eq:Lifshitzmetric}). In the last years, there has
been an intensive activity looking for asymptotically Lifshitz
black holes; some examples are given in
\cite{Taylor:2008tg,AyonBeato:2009nh,AyonBeato:2010tm,Alvarez:2014pra}
and references therein. Holographically, these solutions should
describe the finite temperature behavior of the related
non-relativistic systems. Here $z=1$ describes the isotropic
paradigm since the metric becomes the one of AdS spacetime,
whose stealth configurations were studied in
\cite{Ayon-Beato:SAdS}. Another special situation, although
holographically less motivated, is for $z=0$ since in this case
the resulting Lifshitz background becomes conformally flat.
Nevertheless, as shown below, this last feature induces
nontrivial consequences regarding the existence of conformal
stealths.

More recently, there has been some interest in extending this
kind of metrics (\ref{eq:Lifshitzmetric}) by introducing an
additional parameter, the hyperscaling violation exponent
$\theta$, such that the scaling transformations do not act as
an isometry but rather like a conformal transformation. These
metrics, referred to as hyperscaling violation metrics, are
described by the following line element
\cite{Charmousis:2010zz}
\begin{equation}\label{eq:HSVmetric}
ds_{\mathrm{H}}^2=\left(\frac{l}{r}\right)^{\frac{2\theta}{D-2}}
\left(-\frac{r^{2z}}{l^{2z}}dt^2+\frac{l^2}{r^2}dr^2
+\frac{r^2}{l^2}d\vec{x}^2\right),
\end{equation}
and transform as $ds_{\mathrm{H}}^2\mapsto
\lambda^{\frac{2\theta}{D-2}}ds_{\mathrm{H}}^2$ under the
scaling (\ref{eq:scaling}) supplemented with the holographic
scaling $r\mapsto\lambda^{-1}r$. Notice that this metric is
conformally related to the Lifshitz one
(\ref{eq:Lifshitzmetric}) which is recovered in the limiting
case $\theta=0$. As in the Lifshitz case, there is a physical
interest in looking for black holes whose asymptotic behavior
coincides with the hyperscaling violation metric, see
e.g.~\cite{HyperBH}. For $\theta=D-2$, this spacetime just
becomes the Minkowski one if $z=0$ or $z=1$, whose stealth
configurations were originally reported in
\cite{AyonBeato:2005tu}.

As mentioned before, the first attempt to extend the ideas of
the AdS/CFT correspondence to non-relativistic physics was done
in the context of the symmetry group of the Schr\"odinger
equation for the free particle, which can be viewed as the
non-relativistic cousin of the conformal group. The gravity
dual metric in this case, defining the Schr\"odinger spacetime
\cite{Son:2008ye,Balasubramanian:2008dm}, is given in $D$
dimensions by
\begin{equation}\label{eq:Schrometric}
ds_{\mathrm{S}}^2=\frac{l^2}{y^2}\left[
-\left(\frac{l}{y}\right)^{2(z-1)}du^2-2dudv
+dy^2+d\vec{x}^2\right].
\end{equation}
where $\vec{x}$ is now a $(D-3)$-dimensional vector. A
geometrical derivation of this metric using the conformal
invariance of the Schr\"odinger equation can be found in
\cite{Duval:2008jg}. We have intentionally changed the
notations of the line element to be in perfect accordance with
those in Ref.~\cite{AyonBeato:2006jf,AyonBeato:2005qq} that
will be our guiding principle to derive the general stealth
configuration on the Schr\"odinger background
(\ref{eq:Schrometric}). Indeed, the role of non-relativistic
time in the dynamical scaling (\ref{eq:scaling}) is now played
by the retarded time $u$ while the holographic coordinate $r$
is replaced by $y^{-1}$ such that the boundary is now located
at $y=0$. This class of metrics is invariant not only under the
anisotropic scaling determined by the exponent $z$, but also
under Galilean transformations. For $z=2$, the metric enjoys
the \emph{full} Schr\"odinger symmetry, see for example
\cite{AyonBeato:2011qw} for an account of all these symmetries.
The particular case $z=1$ is maximally symmetric since we again
recover the AdS metric written in light-cone coordinates.
Another case that will deserve a special attention is for
$z=1/2$, since the metric inside the squared brackets becomes
the flat spacetime, i.e.\ the Schr\"odinger background becomes
conformally flat. In contrast with the previous cases, the
presence of the null coordinate $v$ required by the Galileo
boosts, suggests that this metric will possibly describe
non-relativistic quantum theory in dimension $D-2$ and not in
co-dimension one. This remark together with the fact that this
additional coordinate represents a compact direction makes its
holographic interpretation unclear, as previously stated.
However, we shall also study this case in order to be
exhaustive in the comprehension of stealth configurations on
anisotropic backgrounds.

Finally, and consistently with our strategy of completeness, it
is also possible to define a Schr\"odinger inspired background
with hyperscaling violation \cite{Kim:2012nb}. The relevant
line element in this case is given by
\begin{equation}\label{eq:HVSchro}
ds_{\mathrm{HS}}^2=\left(\frac{y}{l}\right)^{\frac{2(\theta-D+2)}{D-2}}
\left[-\left(\frac{l}{y}\right)^{2(z-1)}du^2-2dudv
+dy^2+d\vec{x}^2\right].
\end{equation}
For $\theta=D-2$, this metric describes \emph{pp}-waves, and if
additionally $z=1$ or $z=1/2$ we recover the flat spacetime,
whose stealths were studied in \cite{AyonBeato:2005tu}. These
metrics will end our exploration concerning the existence of
stealths on anisotropic backgrounds.

\section{Stealths on Lifshitz backgrounds}

Here, we will solve the stealth constraints (\ref{eq:SEqs}) on
the Lifshitz background (\ref{eq:Lifshitzmetric}) in full
generality, that is, without any extra assumption. We start by
considering the generic Lifshitz case where the dynamical
critical exponents $z\ne1$ and $z\ne0$ in the first subsection.
In particular, we will establish the holographic nature of the
Lifshitz stealth configuration for those nontrivial values of
the dynamical exponent. The isotropic case, $z=1$, was already
addressed in \cite{Ayon-Beato:SAdS} and the vanishing case,
$z=0$, is poorly motivated from the holographic point of view.
However, for completeness we will also consider this case in
the second subsection.

\subsection{Nontrivial dynamical exponents: Holographic stealths}

For a nonminimal coupling parameter $\xi\ne1/4$, it is useful
to redefine the scalar field as
\begin{equation}\label{eq:Phi2sigma}
\Phi = \frac{1}{\sigma^{2\xi/(1-4\xi)}},
\end{equation}
where $\sigma=\sigma(x^\mu)$ is a local function depending on
all coordinates. The case $\xi=1/4$ will be analyzed at the end
of the subsection. Just for completeness, we mention that
stealth configurations are not allowed in the minimal case,
$\xi=0$, independently of the background. Using the above
redefinition, the off-diagonal components of the
energy-momentum tensor (\ref{eq:eqsTmunu}) along the time
coordinate give rise to the following constraints
\begin{equation}\label{eq:T_t,ri}
T_{\mu{t}}=\frac{(2\xi)^2}{1-4\xi}\left(\frac{r}{l}\right)^z
\frac{\Phi^2}{\sigma}\,
\partial^2_{\mu{t}}\left[\left(\frac{l}{r}\right)^z\sigma\right]=0,
\qquad \mu\neq{t},
\end{equation}
which in turn implies that
\begin{equation}\label{eq:sigmat}
\sigma(t,r,x^i) = \left(\frac{r}{l}\right)^zT(t)+\hat{\sigma}(r,x^i),
\end{equation}
where $T$ (resp. $\hat{\sigma}$) is an arbitrary function of
$t$ (resp. of $r$ and the spatial coordinates). The remaining
off-diagonal constraints are expressed by
\begin{subequations}
\begin{eqnarray}
T_{ri}&=&\frac{(2\xi)^2}{1-4\xi}\frac{r}{l}\frac{\Phi^2}{\sigma}\,
\partial^2_{ri}\left(\frac{l}{r}\hat{\sigma}\right)=0,\label{eq:T_ri}\\
T_{ij}&=&\frac{(2\xi)^2}{1-4\xi}\frac{r}{l}\frac{\Phi^2}{\sigma}\,
\partial^2_{ij}\left(\frac{l}{r}\hat{\sigma}\right)=0, \quad i\neq{j},
\label{eq:T_ij}
\end{eqnarray}
\end{subequations}
and those latter impose that $l\hat{\sigma}/r$ is totally
separable in sum with respect to all its dependencies. This
takes into account all the $D(D-1)/2$ off-diagonal constraints
and permits to conclude that the function $\sigma$ satisfies
the following separability
\begin{equation}\label{eq:sigmasep}
\sigma(t,r,x^i) = \left(\frac{r}{l}\right)^zT(t)+\frac{l}{r}H(r)
+\frac{r}{l}\left[X^1(x^1)+\cdots+X^{D-2}(x^{D-2})\right],
\end{equation}
where $H$ is a function of the holographic coordinate $r$, and
where each function $X^{i}$ only depends on the planar
coordinate $x^i$. Moreover, we emphasize that the functions
involved in the separability are not uniquely defined. In fact,
they are determined modulo the following transformations
\begin{subequations}\label{eq:ResSymm}
\begin{eqnarray}
(X^i, X^j) &\mapsto& (X^i-C^{ij}, X^j+C^{ij}), \\
(X^i, H) &\mapsto& (X^i-C^i, H+C^i\,r^2/l^2), \label{eq:Xi,R}\\
(T, H) &\mapsto& (T-C^0, H+C^0\,r^{z+1}/l^{z+1}), \label{eq:T,R}
\end{eqnarray}
\end{subequations}
where $C^{ij}$, $C^i$ and $C^0$ are arbitrary constants. These
residual symmetries of the separability ansatz
(\ref{eq:sigmasep}) will be useful in the deduction of the
holographic behavior of the stealth.

Let us now consider the diagonal stealth constraints. We start
by analyzing the difference between two different spatial
components
\begin{equation}\label{eq:Tii-Tjj}
T_{(i)}^{~~(i)}-T_{(j)}^{~~(j)}=\frac{(2\xi)^2}{1-4\xi}\frac{l}{r}
\frac{\Phi^2}{\sigma}
\left(\frac{\mathrm{d}^2X^i}{\mathrm{d}(x^i)^2}-
\frac{\mathrm{d}^2X^j}{\mathrm{d}(x^j)^2}\right)=0, \quad i\neq{j},
\end{equation}
where repeated indices between parenthesis mean that there is
no sum for those indices. We infer from these $(D-3)$
requirements that
\begin{equation}\label{eq:alpha}
\frac{\mathrm{d}^2X^1}{\mathrm{d}(x^1)^2}=\cdots=
\frac{\mathrm{d}^2X^i}{\mathrm{d}(x^i)^2}=\cdots=
\frac{\mathrm{d}^2X^{D-2}}{\mathrm{d}(x^{D-2})^2}=\rm{const.}
\end{equation}
Using the above conditions together with
\begin{equation}\label{eq:(Tjj-Trr),i}
\partial_i\left(\frac{\sigma\left(T_{(j)}^{~~(j)}-T_r^{~r}\right)}
{\Phi^2}\right)=
\frac{z(z-1)\xi\,r}{l^3}\frac{\mathrm{d}X^i}{\mathrm{d}x^i}=0,
\end{equation}
we conclude that since $z\ne0$ and $z\ne1$ each function $X^i$
is a constant that can be chosen to be zero without loss of
generality. This can be seen easily by redefining appropriately
the holographic dependence $H(r)$ according to the residual
symmetry (\ref{eq:Xi,R}) of the separability ansatz
(\ref{eq:sigmasep}). Something similar can be deduced for the
temporal dependence from the condition
\begin{equation}\label{eq:(Ttt-Trr),t}
\partial_t\left(\frac{\sigma\left(T_{t}^{~t}-T_r^{~r}\right)}
{\Phi^2}\right)=
\frac{(D-2)(z-1)\,\xi}{l^2}
\frac{\mathrm{d}T}{\mathrm{d}t}\left(\frac{r}{l}\right)^z+
\frac{(2\xi)^2}{(1-4\xi)}\frac{\mathrm{d}^3T}{\mathrm{d}t^3}
\left(\frac{r}{l}\right)^{-z}=0,
\end{equation}
where the coefficients of the different powers of the
holographic coordinate $r$ must vanish independently since the
dynamical exponent is non-vanishing ($z\not=0$). As a direct
consequence, the function $T$ is also a constant whose value
can be taken as zero redefining again the function $H(r)$, but
using now the residual symmetry (\ref{eq:T,R}). Hence, we
conclude that the stealth configuration defined on a Lifshitz
spacetime depends only on the holographic coordinate $r$. We
would like to stress that the situation is clearly different in
the $z=1$ isotropic-relativistic AdS case and in the trivial
case $z=0$ as it can be seen from the expressions
(\ref{eq:(Tjj-Trr),i}) and (\ref{eq:(Ttt-Trr),t}). This makes
evident that the generic nontrivial anisotropy of Lifshitz
spacetimes is the responsible for the holographic behavior of
the stealth scalar field.

The holographic dependence can be additionally fixed by
considering the following stealth constraints
\begin{subequations}\label{eq:holoconst}
\begin{eqnarray}
T_{t}^{~t}-T_{(i)}^{~~(i)}&=&
\frac{(z-1)\,\xi}{(1-4\xi)\,l\,r}\frac{\Phi^2}{\sigma}
\left(4\xi{r}\frac{\mathrm{d}H}{\mathrm{d}r}
+\left[4(z+D-3)\xi-z-D+2\right]H\right)=0,
\label{eq:Ttt-Tii}\\
T_{r}^{~r}-T_{t}^{~t}&=&\frac{\xi}{(1-4\xi)\,l\,r}
\frac{\Phi^2}{\sigma}
\bigg(4\xi{r}^2\frac{\mathrm{d}^2H}{\mathrm{d}r^2}
-4\xi(z+1)\,r\frac{\mathrm{d}H}{\mathrm{d}r}
\nonumber\\
&&\qquad\qquad{}-\left\{4\left[(D-3)z-D+1\right]\xi
-(D-2)(z-1)\right\}H\bigg)=0.
\label{eq:Trr-Ttt}
\end{eqnarray}
\end{subequations}
The first equation imposes a power-law behavior for the
holographic dependence while the second one restricts the value
of the nonminimal coupling parameter $\xi$ in terms of the
dynamical exponent $z$ for any dimension $D$ yielding to
\begin{subequations}\label{eq:solstealth1}
\begin{align}
\xi_{\,\mathrm{L}} &\equiv \frac{(z+D-2)^2}
{4\left[(z+D-2)^2+z^2+D-2\right]},
\label{eq:xi_L}\\
\Phi(r) &= \Phi_0\left(\frac{l}{r}\right)^{\frac{z+D-2}{2}}.
\label{eq:solstealth2}
\end{align}
\end{subequations}
Here, the constant $\Phi_0$ can be tuned arbitrarily using the
scaling symmetry of Lifshitz backgrounds. It remains to
determine the allowed self-interacting potential. However,
evaluating any component of the energy-momentum tensor using
the expressions given by (\ref{eq:solstealth1}) it is easy to
see that $T_{\mu}^{~\nu}=-U(\Phi)\delta_{\mu}^{~\nu}=0$, and
hence only massless free stealth configurations are allowed on
the Lifshitz background.

Many comments can be made concerning this solution. First, it
is interesting to note that the nonminimal coupling parameter
is appropriately parameterized in each dimension $D$ in terms of
the dynamical critical exponent $z$. This situation is quite
different from the stealth solutions on Minkowski or (A)dS
backgrounds since in these cases, the stealths are allowed for
any value of the nonminimal coupling parameter. We also stress
that the anisotropy is directly responsible of this fixing,
since for the $z=1$ isotropic-relativistic AdS case the two
constraints (\ref{eq:holoconst}) reduce to a single one, and
this explains the unrestricted behavior of the nonminimal
coupling parameter for isotropic backgrounds. Finally, we have
seen that in the nontrivial Lifshitz case, also in contrast
with the (A)dS or Minkowski cases, only massless free
configurations are allowed.

Note that from the expression of the allowed nonminimal
coupling parameter (\ref{eq:xi_L}), this latter is bounded from
above as $\xi_{\,\mathrm{L}}<1/4$; that is stealth
configurations are not possible for $\xi>1/4$ and also for the
limiting case $\xi=1/4$. Indeed, in this limiting case, the
appropriate redefinition is given by
$\sqrt{\kappa}\,\Phi(x^\mu)=\exp\left[\sigma(x^\mu)\right]$,
and it is easy to prove along the same lines as before that
$\sigma$ satisfies similar constraints than those in the
generic case implying first the separability
(\ref{eq:sigmasep}) and subsequently the strictly holographic
dependence. We also end with two restrictions similar to those
obtained in (\ref{eq:holoconst}) with the difference that they
are not compatible in the present case.

\subsection{Conformally flat dynamical exponent:
Stealths overflying Lifshitz\label{subsec:Lz=0}}

As was previously emphasized, the value $z=0$ is poorly
motivated from the point of view of holographic applications.
However, it presents some interesting features related with the
observation that Lifshitz spacetime is conformally flat for a
vanishing dynamical exponent, which is manifest after rewriting
the metric according to
\begin{align}
ds_{\mathrm{L}}^2&=-dt^2+\frac{l^2}{r^2}dr^2
+\frac{r^2}{l^2}d\vec{x}^2\nonumber\\
&=\frac{r^2}{l^2}\left\{
-\left[d\left(\frac{l^2}{r}\sinh\frac{t}{l}\right)\right]^2
+\left[d\left(\frac{l^2}{r}\cosh\frac{t}{l}\right)\right]^2
+d\vec{x}^2\right\}
\equiv\Omega^2\eta_{\mu\nu}d\bar{x}^{\mu}d\bar{x}^{\nu},\label{eq:LCF}
\end{align}
where $\{\bar{x}^{\mu}\}$ are the standard cartesian
coordinates of flat spacetime. Moreover, for $z=0$ the Lifshitz
nonminimal coupling (\ref{eq:xi_L}) just becomes the conformal
one, $\xi_{\mathrm{L}}=\xi_D$, defined by
\begin{equation}\label{eq:xi_D}
\xi_D\equiv\frac{D-2}{4(D-1)}.
\end{equation}
Only for the conformal coupling, and for a self-interaction
potential specified as $U(\Phi)\propto\Phi^{\frac{2D}{D-2}}$,
the action (\ref{eq:action}) becomes invariant under conformal
transformations (see Subsec.~\ref{subsec:confmap} for precise
definitions). Any symmetry of the action (\ref{eq:action}) is a
symmetry of the stealth constraints (\ref{eq:SEqs}).
Consequently, for the conformal coupling and the conformal
potential the stealth constraints (\ref{eq:SEqs}) are
conformally invariant and any stealth solution defines a whole
conformal class of stealth solutions. Hence, due to the
conformal flatness of the Lifshitz spacetime for $z=0$, which
is achieved for the conformal factor defined in (\ref{eq:LCF}),
we can build for $\xi=\xi_D$ a conformal stealth on this
background, $\Phi_{\mathrm{L}}$, just by performing a conformal
transformation to the conformal stealth existing on the flat
spacetime, $\Phi_{\mathrm{F}}$, and studied in
\cite{AyonBeato:2005tu}, namely,
\begin{subequations}\label{eq:CSLz=0}
\begin{align}
\Phi_{\mathrm{L}}(x^{\mu}) &= \Omega^{-(D-2)/2}\,
\Phi_{\mathrm{F}}(\bar{x}^{\mu})
 =\left[\Omega\left(\frac{\alpha}2\eta_{\mu\nu}\bar{x}^{\mu}\bar{x}^{\mu}
  +k_{\mu}\bar{x}^{\mu}+\sigma_{0}\right)\right]^{-(D-2)/2}
\nonumber\\
&= \Biggl( \frac{r}{l} \Biggl\{
    \frac{\alpha}2\left[
  - \left(\frac{l^2}{r}\sinh\frac{t}{l}\right)^2
  + \left(\frac{l^2}{r}\cosh\frac{t}{l}\right)^2+\vec{x}^2\right]
  + k_t\left(\frac{l^2}{r}\sinh\frac{t}{l}\right)
  + k_r\left(\frac{l^2}{r}\cosh\frac{t}{l}\right)
  \nonumber\\
&\qquad\quad
  + \vec{k}\cdot\vec{x}
  + \sigma_0           \Biggr\}
   \Biggr)^{-(D-2)/2}\nonumber\\
&= \Biggl\{ lk_t\sinh\frac{t}{l} + lk_r\cosh\frac{t}{l}
  + \frac{r}{l} \Biggl[
    \frac{\alpha}2\left( \frac{l^4}{r^2} + \vec{x}^2 \right)
  + \vec{k}\cdot\vec{x}
  + \sigma_0           \Biggr]
   \Biggr\}^{-(D-2)/2},
\end{align}
where the conformally invariant potential is determined in the
same line as in flat spacetime
\begin{equation}\label{eq:CpotLz=0}
U(\Phi) =\frac{(D-2)^2}{8}\lambda\Phi^{\frac{2D}{D-2}}, \qquad
\lambda=-k_t^2+k_r^2+\vec{k}^2-2\alpha\sigma_0.
\end{equation}
\end{subequations}
This is exactly the same result that is obtained by explicitly
integrating the stealth constraints for $z=0$ and $\xi=\xi_D$.
We found it more useful to present the above argument than the
explicit integration because it makes the origin of this
configuration clear. Here the constant $k_r$ can be eliminated using
the time translation invariance of Lifshitz spacetime, if
$\alpha\ne0$ the same can be done for $\vec{k}$ by means of
space translations and the constant $\alpha$ itself can be
tuned to any value with the help of the scaling symmetry. As a
consequence the solution presents a single independent
integration constant since the other one is determined by the
conformal coupling constant. A similar conclusion is achieved
for $\alpha=0$ using now spatial rotations and the scaling.
This integration constant is a result of conformal symmetry as
it occurs in flat spacetime \cite{AyonBeato:2005tu}.

For $\alpha=0=k_\mu$ we recover the holographic stealth
(\ref{eq:solstealth1}) evaluated at $z=0$. The above conformal
configuration contains in contrast a homogeneous subclass when
$\alpha=0=\vec{k}=\sigma_0$, free of any integration
constants. It is possible to show that this exclusively
time-dependent solution can be generalized to any value of the
nonminimal coupling parameter $\xi$. In fact, for $\xi\ne\xi_D$
no other behavior than the homogeneous one is possible, giving
as result
\begin{subequations}\label{eq:Lsz=0xi}
\begin{align}
\Phi(t) &= \left(\frac{k_t}{\omega}\sinh(\omega{t})
+lk_r\cosh(\omega{t})\right)^{-\frac{2\xi}{1-4\xi}},
&\omega^2 = \frac{(D-2)(1-4\xi)}{4\xi{l}^2},\\
U(\Phi)&=\frac{2\,\xi\Phi^2}{1-4\xi}
\left(\frac{\xi\lambda\Phi^{\frac{1-4\xi}{\xi}}}{1-4\xi}
-\frac{(D-1)\,(D-2)(\xi-\xi_D)}{l^2}
\right),\label{eq:ULsz=0xi}
&\lambda=-k_t^2+l^2\omega^2k_r^2.
\end{align}
\end{subequations}
Notice that for $\xi=\xi_D$ we consistently recover the
homogeneous version of the conformal solution
(\ref{eq:CSLz=0}).

The case $\xi=1/4$ is excluded from the previous analysis since
it is a priori not covered by the redefinition
(\ref{eq:Phi2sigma}). However, it is possible to show the
existence of a stealth solution in this case that remarkably
can be obtained as a nontrivial limit of the above solution. We
shall explain the involved procedure in a general setting,
since it has the potential to be applied in other contexts. We
start reconsidering the redefinition (\ref{eq:Phi2sigma}) by
introducing explicitly the dimension of the scalar field as
$\sqrt{\kappa}\,\Phi = \sigma^{-2\xi/(1-4\xi)}$, being $\kappa$
the Einstein constant, in order to have a dimensionless
redefined function $\sigma$. The solution for $\sigma$ will be
exactly the same, with the difference that the integration
constants must have now the proper dimensions that make
$\sigma$ dimensionless; in the present case the constants $k_t$
and $k_r$ in (\ref{eq:Lsz=0xi}) will change to have dimensions
of inverse length. The remaining procedure is simple, if after
possibly redefining the integration constants we find that the
following limit is well-behaved
\begin{equation}\label{eq:hatsigma}
\lim_{\xi\rightarrow1/4}\frac{2\xi(1-\sigma(x^\mu))}
{1-4\xi}\equiv\hat{\sigma}(x^\mu),
\end{equation}
the configuration for $\xi=1/4$ can be obtained as
\begin{align}
\Phi &= \lim_{\xi\rightarrow1/4}
\frac1{\sqrt{\kappa}}\,\sigma^{-2\xi/(1-4\xi)}\nonumber\\
&=\lim_{\xi\rightarrow1/4}
\frac1{\sqrt{\kappa}}\,
\left(1-\frac{1-4\xi}{2\xi}\hat{\sigma}+O\left((1-4\xi)^2\right)\right)^{-2\xi/(1-4\xi)}
\nonumber\\
&= \frac1{\sqrt{\kappa}}\,e^{\hat{\sigma}},\label{eq:xi->1/4exp}
\end{align}
where we have used the limit definition of the exponential
function $e^x\equiv\lim_{m\rightarrow0}(1+mx)^{1/m}$. For the
solution (\ref{eq:Lsz=0xi}) the condition (\ref{eq:hatsigma})
is achieved after redefining the integration constants by
\begin{equation}\label{eq:redk2h}
k_t=-\frac{1-4\xi}{2\xi}\hat{k}_t, \qquad
k_r=\frac1l\left(1-\frac{1-4\xi}{2\xi}\hat{\sigma}_0\right),
\end{equation}
which gives
\begin{equation}\label{eq:hsLz=0}
\hat{\sigma}(t)=-\frac{D-2}4\frac{t^2}{l^2}+\hat{k}_tt+\hat{\sigma}_0.
\end{equation}
This procedure must be also consistent when applied to the
supporting self-interactions, i.e.\ taking the nontrivial limit
must produce a well-behaved result. For the potential
(\ref{eq:ULsz=0xi}) the limit resulting from the redefinitions
(\ref{eq:redk2h}), after considering that the coupling constant
is now related to the integration constants by
$\lambda=\kappa^{(1-4\xi)/(2\xi)}(-k_t^2+l^2\omega^2k_r^2)$, is
the following
\begin{align}
U(\Phi) &= \lim_{\xi\rightarrow1/4}
\frac{(D-2)\Phi^2}{8l^2}
\left\{\frac{4\xi}{1-4\xi}
\left[\left(\frac{\Phi}{\Phi_0}\right)^{(1-4\xi)/\xi}-1\right]
+D-1 +O\left(1-4\xi\right)\right\}
\nonumber\\
&= \frac{(D-2)\Phi^2}{8l^2}
\left[4\ln\left(\frac{\Phi}{\Phi_0}\right)+D-1\right],\label{eq:xi->1/4U}
\end{align}
where the coupling constant is given by
$\sqrt{\kappa}\,\Phi_0=\exp[{l^2\hat{k}_t^2}/{(D-2)}+\hat{\sigma}_0]$
and additionally we have used the limit definition of the
logarithmic function
$\ln{x}\equiv\lim_{m\rightarrow0}(x^m-1)/m$. The configuration
(\ref{eq:xi->1/4exp}), (\ref{eq:hsLz=0}) with the
self-interaction (\ref{eq:xi->1/4U}) is precisely the stealth
solution resulting from integrating explicitly the stealth
constraints for $\xi=1/4$ when Lifshitz spacetime has a
vanishing dynamical exponent. This exhausts all the
nonminimally coupled scalar stealths allowed to exist on the
Lifshitz background. In the next section we continue with a
similar search on hyperscaling violating backgrounds.

\section{Stealths in presence of hyperscaling violation}

We now look for the existence of stealth configurations defined
on hyperscaling violation metric (\ref{eq:HSVmetric}). Since
this background is conformally related to the Lifshitz
spacetime (\ref{eq:Lifshitzmetric}), we first use a conformal
argument in order to obtain hyperscaling violation stealth
solutions from those derived in the Lifshitz case in
Subsec.~\ref{subsec:confmap}. We will see later that the
configuration obtained through the conformal mapping represents
only a particular class of the stealth configurations on
hyperscaling violation metric by deriving the most general
solution in Subsec.~$4.2$. The cases associated to the
nonminimal coupling $\xi=1/4$ are separately analyzed in
Subsec.~$4.3$.

\subsection{\label{subsec:confmap}Holographic branch:
Conformal map from Lifshitz}

In the standard AdS/CFT correspondence, it is well-known that
the AdS metric solves the Einstein equations with a negative
cosmological constant. In contrast, in order to support
Lifshitz spacetimes (\ref{eq:Lifshitzmetric}) the vacuum
Einstein equations are not enough and they require the
introduction of some matter source,
e.g.~\cite{Taylor:2008tg,Alvarez:2014pra}, or a more radical
approach consists in considering higher order gravity theories
as done in Refs.~\cite{AyonBeato:2009nh,AyonBeato:2010tm}. In
the case of the hyperscaling violation metric
(\ref{eq:HSVmetric}), a simple computation shows that this
spacetime is solution of the vacuum Einstein equations
\cite{Dong:2012se}, without cosmological constant, provided the
dynamical and hyperscaling violation exponents are fixed in
term of the dimension $D$ as
\begin{equation}\label{eq:solGR}
z=\frac{2(D-2)}{D-3},\qquad
\theta=\frac{(D-1)(D-2)}{D-3}.
\end{equation}
Here, we first re-derive this result by an indirect method. In
fact, we will establish a correspondence between the
holographic stealth configuration (\ref{eq:solstealth1}) on the
Lifshitz spacetime (\ref{eq:Lifshitzmetric}) defined by a free
massless nonminimally coupled scalar field with a solution of
the vacuum Einstein equations in the specific case where the
nonminimal coupling is given by the conformal one $\xi_D$
defined in (\ref{eq:xi_D}). Indeed, it is well-known that the
action (\ref{eq:action}) with $\xi=\xi_D$ and without potential
is conformally invariant. More precisely, under the conformal
transformation
\begin{equation}
\bar{g}_{\mu\nu}=\Omega^2\,g_{\mu\nu},\qquad
\bar{\Phi}=\Omega^{-(D-2)/2}\,\Phi,
\label{eq:conftransf}
\end{equation}
where the conformal factor $\Omega=\Omega(x)$ is any local
function, the action is invariant up to a boundary term
\begin{equation}\label{eq:confinv}
S_{\xi_D}^0[\Phi,g_{\mu\nu}]=S_{\xi_D}^0[\bar{\Phi},\bar{g}_{\mu\nu}]
+\rm{b.t.},
\end{equation}
where we denote as $S_{\xi}^0$ the part of action
(\ref{eq:action}) without the self-interaction contribution. A
consequence of this symmetry is that in the conformal frame
defined by the particular choice of the conformal factor
$\Omega=(\Phi/\Phi_0)^{2/(D-2)}$, the scalar field turns into a
constant value $\Phi_0$ and the action becomes proportional to
the Einstein-Hilbert action. At the level of the field
equations, this translates to the fact that solutions of the
stealth equations (\ref{eq:SEqs}) in the case of the conformal
coupling (\ref{eq:xi_D}) map to solutions of the vacuum
Einstein equations. As we have shown in the previous section,
the general solution of the stealth constraints (\ref{eq:SEqs})
defined on a Lifshitz spacetime (\ref{eq:Lifshitzmetric}) with
nontrivial dynamical exponent $z$ is given by
(\ref{eq:solstealth1}). Then, it is easy to realize that for
the dynamical critical exponent $z=2(D-2)/(D-3)$, the Lifshitz
stealth becomes a conformally invariant configuration since in
this case $\xi_{\,\mathrm{L}}=\xi_D$. Additionally, at the
mentioned conformal frame the vacuum metric conformally related
to the Lifshitz one exhibits hyperscaling violation with
exponent $\theta=(D-1)(D-2)/(D-3)$. Hence, we have justified
the existence of General Relativity vacua for the hyperscaling
violation metric with exponents given by (\ref{eq:solGR}), but
using a very simple and elegant conformal argument. Note that
the holographic character of the Lifshitz stealth is behind
this mapping. Indeed, if the scalar field would have depended
on some other coordinates, this mapping to the hyperscaling
violation metric would not have been possible.

A less known fact is the change of the action (\ref{eq:action})
for a non-conformal coupling $\xi\neq\xi_D$ under a conformal
transformation (\ref{eq:conftransf}). Actually, in a
generalized conformal frame defined by
$\Omega=(\Phi/\Phi_0)^\alpha$, remarkably, the action without
potential becomes proportional to the same action but with a
different nonminimal coupling. More precisely, the following
$1$-parameter field transformations
\begin{subequations}\label{eq:map}
\begin{equation}
\bar{g}_{\mu\nu}=\left(\frac{\Phi}{\Phi_0}\right)^{2\alpha}g_{\mu\nu},\qquad
\bar{\Phi}=\Phi_0\left(\frac{\Phi}{\Phi_0}\right)^{[2-(D-2)\alpha]/2},\\
\end{equation}
together with the reparameterization
\begin{equation}\label{eq:barxi}
\bar{\xi}=\frac14\frac{\left[(D-2)\alpha-2\right]^2\xi}{1+\alpha\xi(D-1)
\left[(D-2)\alpha-4\right]},
\end{equation}
\end{subequations}
define a map between the actions of two massless free scalar
fields with nonminimal couplings given by $\xi$ and
$\bar{\xi}$. Concretely, we have
\begin{equation}\label{eq:actionmap}
\frac1{\xi}\,S_{\xi}^0[\Phi,g_{\mu\nu}]=
\frac1{\bar{\xi}}\,S_{\bar{\xi}}^0[\bar{\Phi},\bar{g}_{\mu\nu}]+\rm{b.t.}
\end{equation}
Notice that when both nonminimal couplings coincide the
transformation (\ref{eq:map}) obviously becomes a symmetry of
the nonminimally coupled action, which is nothing but the
conformal symmetry since this occurs only if
$\bar{\xi}=\xi=\xi_D$.

We will exploit the above conformal argument to build a stealth
configuration on the hyperscaling violation metrics
(\ref{eq:HSVmetric}) using as seed those existing on the
Lifshitz background (\ref{eq:Lifshitzmetric}) and defined by
(\ref{eq:solstealth1}). The generalized conformal frame defined
by $\Omega=(\Phi/\Phi_0)^\alpha$ will correspond to the
hyperscaling violation metric (\ref{eq:HSVmetric}) only if
$\alpha=2\theta/[(D-2)(z+D-2)]$. This in turn implies that the
configuration given by
\begin{subequations}\label{eq:stealthH}
\begin{align}
\xi_{\,\mathrm{H}}&\equiv\frac14\frac{(D-2)(\theta-z-D+2)^2}
{(D-1)(\theta-z-D+2)^2 +(D-3)z^2-2(D-2)z},\label{eq:xi_H}\\
\Phi(r)&=\Phi_0\left(\frac{l}{r}\right)^{\frac{z+D-2-\theta}2},
\label{eq:stealthHYsf}
\end{align}
\end{subequations}
defines a stealth on the hyperscaling violation metric
(\ref{eq:HSVmetric}). Consistently, the limit $\theta\to0$
reduces to the stealth solution on the Lifshitz background
(\ref{eq:solstealth1}). We can also observe that for the values
of the exponents which solve the vacuum Einstein equations
(\ref{eq:solGR}), the stealth scalar field
(\ref{eq:stealthHYsf}) becomes precisely a constant and the
hyperscaling violating nonminimal coupling (\ref{eq:xi_H})
becomes just the conformal one (\ref{eq:xi_D}). In fact,
$\xi_{\,\mathrm{H}}=\xi_D$, only if the dynamical exponent
takes the value (\ref{eq:solGR}) analyzed at the beginning of
the subsection or if it vanishes. This last case will be
studied at the end of the next subsection since it entails more
general configurations than the holographic ones, due to the
emergence of conformal symmetry for $z=0$.

\subsection{Non-holographic branches: Full derivation}

The conformal argument establishing the existence of stealth
configurations for spacetimes with hyperscaling violation is
undoubtedly a very elegant one. Unfortunately, it lacks the
skill to exclude the existence of more general configurations,
in particular non-holographic ones. Before, we just establish
that for a generic value of the hyperscaling violation exponent
$\theta$ a stealth solution can be conformally generated, which
is holographic by construction. However, as shown below there
also exist two special families for particular values for
$\theta$, parameterized in terms of the dynamical critical
exponent $z$ for any dimension, which give rise to stealth
configurations that are not holographic. These latter are
additionally self-interacting in contrast to the free behavior
of the holographic one obtained from the previous conformal
mapping. Moreover, as in the Lifshitz situation, the
conformally flat case $z=0$ deserves a separate analysis that
we will perform at the end of the subsection. Hence, most of
this subsection is restricted to the study of anisotropic
backgrounds with nontrivial dynamical exponents, $z\ne1$ and
$z\ne0$.

In order to make transparent the previously anticipated
conclusion we shall make use of two nontrivial linear
combinations of the stealth constraints (\ref{eq:SEqs})
together with their derivatives, all appropriately evaluated on
the hyperscaling violating spacetime (\ref{eq:HSVmetric}). We
start studying the off-diagonal stealth constraints, where we
make use again of the redefinition (\ref{eq:Phi2sigma}), and we
yield to the following separation
\begin{equation}\label{eq:sigmasepH}
\sigma(t,r,x^i) = \left(\frac{l}r\right)^{\frac{\theta}{D-2}}
\left[\left(\frac{r}{l}\right)^zT(t)+\frac{l}{r}H(r)
+\frac{r}{l}\left[X^1(x^1)+\cdots+X^{D-2}(x^{D-2})\right]\right],
\end{equation}
where the unknown functions are again defined modulo the same
residual symmetries characterizing the Lifshitz case
(\ref{eq:ResSymm}). Additionally, the differences between the
diagonal spatial components once more establish  the conditions
(\ref{eq:alpha}). The first nontrivial combination we now use
is given by {\small
\begin{align}\label{eq:ntlc1H}
\left(\frac{l}r\right)^{\frac{\theta}{D-2}}\frac{\sigma}{\Phi^2}\left[
\frac{(D-1)(\theta-z-D+2)(\xi-\xi_D)+z\,\xi}
{(D-2)\xi}\left(T_t^{~t}-T_{(i)}^{~~(i)}\right)
-(z-1)\left(T_r^{~r}-T_{(i)}^{~~(i)}\right)\right]&\nonumber\\\nonumber\\
{}+r\partial_r\left[\left(\frac{l}r\right)^{\frac{\theta}{D-2}}
\frac{\sigma}{\Phi^2}\left(T_t^{~t}-T_{(i)}^{~~(i)}\right)\right]
=\frac{(z-1)(\theta-z-D+2)^2(\xi-\xi_{\,\mathrm{H}})}
{4l^2\xi_{\,\mathrm{H}}}
\left(\frac{r}l\right)^{\frac{\theta}{D-2}}\sigma&\nonumber\\\nonumber\\
{}-\frac{4\xi\left[(D-1)(\theta-z-D+2)(\xi-\xi_D)-(D-3)z\,\xi\right]}
{(D-2)(1-4\xi)}
\left[\frac{\mathrm{d}^2T}{\mathrm{d}t^2}\left(\frac{r}l\right)^{-z}
+\frac{\mathrm{d}^2X^i}{\mathrm{d}(x^i)^2}\frac{l}r\right]&=0.
\end{align}}%
The first conclusion we can draw from here is that not only in
the holographic case the nonminimal coupling is restricted to
take the value $\xi=\xi_{\,\mathrm{H}}$, but also this
restriction must be satisfied for any potentially
non-holographic configuration. This can be easily viewed by
taking the derivative with respect to any non-holographic
coordinate of the right-hand side of (\ref{eq:ntlc1H}). Notice
that these conclusions would be different if $z=1$ or $z=0$,
and this the reason why these cases deserve a separate
attention. We now consider the following second nontrivial
combination evaluated at $\xi=\xi_{\,\mathrm{H}}$ {\small
\begin{align}\label{eq:ntlc2H}
\frac{(1-4\,\xi_{\,\mathrm{H}})}{4{\xi_{\,\mathrm{H}}}^2}
\left(\frac{l}r\right)^{\frac{\theta}{D-2}}\frac{\sigma}{\Phi^2}
\left((z-1)T_r^{~r}-zT_t^{~t}+T_{(i)}^{~~(i)}\right)&\nonumber\\\nonumber\\
=\frac{z(z-1)\{[(D-3)z-D+2]\theta-(D-2)^2(z-1)\}[\theta-(D-2)(z-1)]}
{(D-2)^2(\theta-z-D+2)^2l^2}
\left(\frac{r}l\right)^{\frac{\theta}{D-2}}\sigma&\nonumber\\\nonumber\\
{}+z\frac{\mathrm{d}^2T}{\mathrm{d}t^2}\left(\frac{r}l\right)^{-z}
+\frac{\mathrm{d}^2X^i}{\mathrm{d}(x^i)^2}\frac{l}r
+\frac{z-1}{lr}\left(r\frac{\mathrm{d}~}{\mathrm{d}r}-2\right)\!\!
\left(r\frac{\mathrm{d}~}{\mathrm{d}r}-(z+1)\right)\!\!H
&=0.
\end{align}}%
Again, taking the derivative with respect to any
non-holographic coordinate, we arrive to the conclusion that
non-holographic stealth configurations are only allowed for the
following two values of the hyperscaling violating exponent
\begin{subequations}\label{eq:thetasNH}
\begin{align}
\theta&=\frac{(D-2)^2(z-1)}{(D-3)z-D+2},\label{eq:thetaf}\\
\theta&=(D-2)(z-1).\label{eq:thetas}
\end{align}
\end{subequations}
One also concludes that not only the second derivatives of the
functions holding the spatial dependence are constant
(\ref{eq:alpha}) but also the second derivative of the function
that encloses the time evolution. Returning to the former
restriction (\ref{eq:ntlc1H}) and considering that
$\xi=\xi_{\,\mathrm{H}}$ together with the previous values for
the hyperscaling violating exponent allowing non-holographic
behaviors, it is easy to realize that the coefficient in front
of these constant second derivatives at (\ref{eq:ntlc1H}) never
vanishes; hence the second derivatives of the functions
enclosing the non-holographic dependencies are all necessarily
zero, which entails only linear non-holographic regimes for the
redefinition of the stealth scalar field (\ref{eq:Phi2sigma}).

For hyperscaling violating exponents different from the values
(\ref{eq:thetasNH}) the stealth must be necessarily
holographic. An exhaustive study of those cases just give rises
to the stealth conformally constructed from the Lifshitz one
and derived in the previous subsection (\ref{eq:stealthH}).

The hyperscaling violating exponents (\ref{eq:thetasNH}) are
mutually exclusive in an anisotropic context ($z\ne1$) since
the resulting nonminimal coupling (\ref{eq:xi_H}) must be
nontrivial in order for any stealth configuration to be
possible; in fact, the vacuum related exponents
(\ref{eq:solGR}) are the only anisotropic possibility.
Actually, the exponents (\ref{eq:thetasNH}) are not only
different numerically but, as discussed below, they also
characterize qualitatively different non-holographic behaviors.

Let us start analyzing the first case where the hyperscaling
violating exponent takes the value (\ref{eq:thetaf}).
Evaluating the next combination at this value we obtain
\begin{equation}
\frac{(1-4\,\xi_{\,\mathrm{H}})}{4{\xi_{\,\mathrm{H}}}^2}
\left(\frac{l}r\right)^{\frac{\theta}{D-2}}\frac{\sigma}{\Phi^2}
\left(T_t^{~t}-T_{(i)}^{~~(i)}\right)
=\frac{z-1}{l^2}\bigg[\frac{l}r\left(r\frac{\mathrm{d}~}{\mathrm{d}r}
-(z+1)\!\right)H-(z-1)\frac{r}l\sum_{j=1}^{D-2}X^j\bigg]=0.
\label{eq:Ttt-Tii_theta_o}
\end{equation}
Taking the derivative with respect to the spatial coordinates
of the previous combination it is possible to conclude that the
spatial functions are all constants, which can be taken to be
zero using the residual symmetry (\ref{eq:Xi,R}) of the
separability ansatz (\ref{eq:sigmasepH}). The resulting
equation easily fixes the holographic dependence and since it
is just a first integral of Eq.~(\ref{eq:ntlc2H}) this last one
is now completely satisfied. All the stealth constraints are
satisfied except the one that fixes the self-interaction which
turns to be a power law of the scalar field. Finally, the first
class of non-holographic stealth configurations defined in the
hyperscaling violation metric is given by
\begin{subequations}\label{eq:stealthHo}
\begin{align}
\theta&=\frac{(D-2)^2(z-1)}{(D-3)z-D+2},\label{eq:theta_o}\\
\xi_{\,\mathrm{Ho}}&\equiv\frac{z[(D-3)z-2(D-2)]}{4[2(D-3)z^2-4(D-2)z+D-2]},
\label{eq:xi_Ho}\\
\Phi(x^\mu)&=\left[
\left(\frac{l}r\right)^{\frac{z(D-2)-\theta}{D-2}}
\frac1{k_tt+\sigma_0}\right]^{\frac{2\,\xi_{\,\mathrm{Ho}}}
{1-4\,\xi_{\,\mathrm{Ho}}}},
\label{eq:Phi(x)Ho}\\
U(\Phi)&=\frac{2\,{\xi_{\,\mathrm{Ho}}}^2}{(1-4\,\xi_{\,\mathrm{Ho}})^2}
\,\lambda\,\Phi^{(1-2\,\xi_{\,\mathrm{Ho}})/\xi_{\,\mathrm{Ho}}},
&\lambda=-k_t^2.
\label{eq:Uo}
\end{align}
\end{subequations}
Note that for $k_t=0$, we just recover the holographic stealth
(\ref{eq:stealthH}) for the hyperscaling violating exponent
(\ref{eq:theta_o}). For $k_t\neq0$, the constant $\sigma_0$ can
be put to zero by a time translation due to the stationarity of
the hyperscaling violating metric (\ref{eq:HSVmetric}). We
emphasize that the self-interaction (\ref{eq:Uo}) is negative
definite since the involved coupling constant $\lambda$ must be
strictly negative. This overflying stealth is not characterized
by any free integration constant.

Let us now consider the other option characterized by a
hyperscaling violating exponent with value (\ref{eq:thetas}).
In this case the same combination after being evaluated at this
exponent gives
\begin{equation}
\frac{(1-4\,\xi_{\,\mathrm{H}})}{4{\xi_{\,\mathrm{H}}}^2}
\left(\frac{l}r\right)^{\frac{\theta}{D-2}}\frac{\sigma}{\Phi^2}
\left(T_t^{~t}-T_{(i)}^{~~(i)}\right)
=\frac{z-1}{l^2}\bigg[\frac{l}r\left(r\frac{\mathrm{d}~}{\mathrm{d}r}
-2\right)H+(z-1)\left(\frac{r}l\right)^zT\bigg]=0.
\label{eq:Ttt-Tii_theta_i}
\end{equation}
We conclude that the temporal dependence is constant and can be
chosen to be zero using the residual symmetry (\ref{eq:T,R}).
From here, we find the holographic dependence, which is also
compatible with Eq.~(\ref{eq:ntlc2H}) since
(\ref{eq:Ttt-Tii_theta_i}) becomes again its first integral.
Finally, along the same lines as before, the second
non-holographic configuration reads
\begin{subequations}\label{eq:stealthHi}
\begin{align}
\theta&=(D-2)(z-1),\frac{}{}\label{eq:theta_i}\\
\xi_{\,\mathrm{Hi}}&\equiv\frac{(D-3)z-2(D-2)}{4[(D-2)z-2(D-1)]},
\label{eq:xi_Hi}\\
\Phi(x^\mu)&=\left[\left(\frac{r}l\right)^{z-2}
\frac1{\vec{k}\cdot\vec{x}+\sigma_0}\right]^{\frac{2\,\xi_{\,\mathrm{Hi}}}
{1-4\,\xi_{\,\mathrm{Hi}}}},
\label{eq:Phi(x)Hi}\\
U(\Phi)&=\frac{2\,{\xi_{\,\mathrm{Hi}}}^2}{(1-4\,\xi_{\,\mathrm{Hi}})^2}
\,\lambda\,\Phi^{(1-2\,\xi_{\,\mathrm{Hi}})/\xi_{\,\mathrm{Hi}}},
&\lambda=\vec{k}^2.\label{eq:Ui}
\end{align}
\end{subequations}
Note that if $\vec{k}=0$, this solution becomes the holographic
stealth (\ref{eq:stealthH}) when the hyperscaling violating
exponent is given by (\ref{eq:theta_i}). Instead, if
$\vec{k}\neq0$, the constant $\sigma_0$ can be assumed again as
vanishing using the invariance under spatial translation
exhibited by the hyperscaling violating metric
(\ref{eq:HSVmetric}). This metric is additionally invariant
under spatial rotation, which allows to fix the inhomogeneity
along any preferred direction eliminating all the components of
the vector $\vec{k}$ except one, which is related to the
coupling constant $\lambda$ (\ref{eq:Ui}). Hence, no
independent integration constant characterizes this
inhomogeneous stealth as in the previous case but with the
difference that now the self-interaction is positive definite.

As was emphasized in this analysis, the above conclusions are
valid in the anisotropic case $z\ne1$ and for nonvanishing
dynamical exponents $z\ne0$. We end this subsection by
examining what is special about the case $z=0$. For Lifshitz
backgrounds this case was considered in
Subsec.~\ref{subsec:Lz=0} and represents a conformally flat
spacetime. Due to the conformal relation with the Lifshitz
backgrounds which defines the present spacetimes
(\ref{eq:HSVmetric}) the vanishing dynamical exponent also
characterizes here a conformally flat spacetime. Additionally,
for $z=0$ the hyperscaling violating nonminimal coupling
(\ref{eq:xi_H}) just becomes $\xi_{\,\mathrm{H}}=\xi_D$, i.e.\
the conformal coupling (\ref{eq:xi_D}). Contrary to the cases
previously studied in this subsection the holographic behavior
is lost now for any value of the hyperscaling violating
exponent $\theta$, which induces a self-interaction which is
just the conformal one. The emergence of conformal symmetry
here also, implies that the corresponding stealth is just a
conformal transformation of the conformal stealth of the
Lifshitz spacetime (\ref{eq:CSLz=0})
\begin{equation}
\Phi_{\mathrm{H}}=\left(\frac{r}{l}\right)^{\frac\theta2}
\Phi_{\mathrm{L}},
\end{equation}
with a conformal potential whose coupling constant is specified
again as in (\ref{eq:CpotLz=0}). This is the only nontrivial
situation allowed for $z=0$. We emphasize that when additionally
$\theta=D-2$ this background becomes precisely flat spacetime,
whose stealths were studied in Ref.~\cite{AyonBeato:2005tu} and
exist for any value of the nonminimal coupling parameter.
Finally, we mention by completeness that other case not studied
here is $z=1$, this case also represents a conformally flat
spacetime since the background is conformal to AdS, hence a
conformally generated solution as the previous one also exists.
The isotropy of this case makes it more rich since it
additionally allows a different class of holographic
configurations valid for any value of the nonminimal coupling,
we do not bring the related details here because the emphasis
of this work is in anisotropy.

\subsection{Nonminimal coupling $\xi=1/4$}

We end this section by analyzing the case $\xi=1/4$ which is
outside of the previous derivation since we have assumed the
redefinition (\ref{eq:Phi2sigma}). In the present situation, it
is more pertinent to redefine the scalar field as
\begin{equation}\label{eq:Phi2sigmaxi1/4}
\Phi = \frac1{\sqrt{\kappa}}\,\mathrm{e}^{\sigma}.
\end{equation}
First, notice that contrary to the Lifshitz nonminimal coupling
(\ref{eq:xi_L}), the hyperscaling violating nonminimal coupling
(\ref{eq:xi_H}) is not bounded from above; hence, in principle,
it can achieve the value $\xi_{\,\mathrm{H}}=1/4$. In fact, a
careful study of holographic stealths with coupling $\xi=1/4$
on these spacetimes just reproduces the stealth
(\ref{eq:stealthH}) conformally generated from the Lifshitz one
when $\xi_{\,\mathrm{H}}=1/4$, which is obviously obtained for
a couple of values of the exponent $\theta$ satisfying a
resulting quadratic equation.

Something similar occurs with the nonminimal couplings
$\xi_{\,\mathrm{Ho}}$ and $\xi_{\,\mathrm{Hi}}$ allowing the
non-holographic behaviors (\ref{eq:stealthHo}) and
(\ref{eq:stealthHi}), respectively, since both are compatible
with the hyperscaling violating nonminimal coupling
(\ref{eq:xi_H}) and consequently can reach the value $1/4$.
The related configurations can be obtained from
(\ref{eq:stealthHo}) and (\ref{eq:stealthHi}) via the
nontrivial limit outlined at the end of
Subsec.~\ref{subsec:Lz=0}. In order to obtain well-behaved
limits we need to redefine the involved integration constants
according to
\begin{equation}\label{eq:redks2hkP0}
k_\mu=-\frac{1-4\xi}{2\xi}\hat{k}_\mu, \qquad
\sigma_0=1-\frac{1-4\xi}{2\xi}\ln\left(\sqrt{\kappa}\,\Phi_0\right),
\end{equation}
and use the appropriated value of the nonminimal coupling in
each case.

Starting from the first example (\ref{eq:stealthHo}) the value
$\xi_{\,\mathrm{Ho}}=1/4$ is achieved for the two dynamical
exponents $z_{\pm}$ defined below, which reduce the
hyperscaling violating exponent of the solution to
$\theta=(D-2)z_{\pm}$. Taking the nontrivial limit
$z{\rightarrow}z_{\pm}$ in (\ref{eq:stealthHo}), after applying
the redefinitions (\ref{eq:redks2hkP0}) evaluated at
$\xi=\xi_{\,\mathrm{Ho}}$, we obtain the following
configuration which describes a massive free stealth overflying
the spacetime
\begin{subequations}\label{eq:stealthHoxi1/4}
\begin{align}
z_{\pm}&=\frac{\sqrt{D-2}}{D-3}\left(\sqrt{D-2}\pm1\right),\\
\theta&=(D-2)z_{\pm},\frac{}{}\label{eq:theta_oxi1/4}\\
\Phi(x^\mu)&=\Phi_0\,\mathrm{e}^{\hat{k}_tt}\left(\frac{r}l\right)
^{\pm\frac{\sqrt{D-2}}2},\label{eq:Phi(x)Hoxi1/4}\\
U(\Phi)&=\frac12\,m^2\,\Phi^2,
&m^2=-\hat{k}_t^2.\label{eq:Uoxi1/4}
\end{align}
\end{subequations}
For $\hat{k}_t=0$, the above configuration reduces to the
holographic stealth (\ref{eq:stealthH}) when the exponents take
the previous values. If $\hat{k}_t\neq0$, the constant $\Phi_0$
can be tuned using time translations, and the free fields have
a tachyonic behavior. Again, these stealths are free of any
independent integration constant.

Finally, starting from the second example (\ref{eq:stealthHi})
the coupling becomes $\xi_{\,\mathrm{Hi}}=1/4$ only if $z=2$,
this implies that the other exponent must take the value
$\theta=D-2$. Now considering the nontrivial limit
$z{\rightarrow}2$ in (\ref{eq:stealthHi}), after using the
redefinitions (\ref{eq:redks2hkP0}) for
$\xi=\xi_{\,\mathrm{Hi}}$, we obtain the following
inhomogeneous massive configurations
\begin{subequations}\label{eq:stealthHixi1/4}
\begin{align}
z&=2,\frac{}{}\\
\theta&=D-2,\frac{}{}\label{eq:theta_ixi1/4}\\
\Phi(x^\mu)&=
\Phi_0\,\frac{l}r\exp{\!\left(\vec{\hat{k}}\cdot\vec{x}\right)},
\label{eq:Phi(x)Hixi1/4}\\
U(\Phi)&=
\frac12\,m^2\,\Phi^2, & m^2=\vec{\hat{k}}^2. \label{eq:Uixi1/4}
\end{align}
\end{subequations}
Once more, if $\vec{\hat{k}}=0$ the configuration becomes the
holographic stealth (\ref{eq:stealthH}) for the present
exponents while for $\vec{\hat{k}}\neq0$ we fix the constant
$\Phi_0$ by translations, the direction of the vector
$\vec{\hat{k}}$ by rotations and its length by the mass $m$.
Consequently, no independent integration constant characterizes
this stealth.

All these are precisely the unique solutions which are obtained
by straightforwardly integrating the stealth constraints
(\ref{eq:SEqs}) on the hyperscaling violating background
(\ref{eq:HSVmetric}) for the nonminimal coupling $\xi=1/4$.

\section{Stealths on Schr\"odinger backgrounds}

In this section we analyze the consequences on the existence of
stealths for the spacetime realization of anisotropic scaling,
without and with hyperscaling violation, but this time using as
gravitational duals the backgrounds (\ref{eq:Schrometric}) and
(\ref{eq:HVSchro}), inspired by the symmetries of the
Schr\"odinger equation for a free particle. Since the methods
are similar to those used in the Lifshitz cases, here we will
only mention and discuss the different allowed cases without
including their deduction.

\subsection{Schr\"odinger inspired dynamical scaling}

Following similar arguments to those described in the Lifshitz
case and using the results obtained in
\cite{AyonBeato:2006jf,AyonBeato:2005qq}, we can derive the
most general solutions of the stealth constraints
(\ref{eq:SEqs}) on a Schr\"odinger background
(\ref{eq:Schrometric}). First, for a nontrivial value of the
dynamical critical exponent $z\not=1$ and $z\not=0$ it can be
concluded again that stealth configurations are only possible
on the Schr\"odinger background for a precise nonminimal
coupling parameter
\begin{subequations}\label{eq:gSS}
\begin{equation}
\xi_{\mathrm{S}}\equiv\frac{2z+D-3}{4(2z+D-2)}.
\label{xiSchrod}
\end{equation}
The isotropic case $z=1$ once again corresponds to AdS space
whose stealths were studied in \cite{Ayon-Beato:SAdS} and the
vanishing case $z=0$ will be treated separately at the end of
the subsection. For a generic value of the exponent the stealth
configuration is self-interacting and determined by
\begin{align}
\Phi(x^\mu)&=\left(\frac{y}{l}\frac1{k_uu+\vec{k}{\cdot}\vec{x}+\sigma_0}
\right)^{\frac{2z+D-3}{2}},
\label{PhiSchro}\\\nonumber\\
U(\Phi)&=\frac{2\,\xi_{\mathrm{S}}\Phi^2}{(1-4\xi_{\mathrm{S}})^2}
\left(\xi_{\mathrm{S}}\lambda
\Phi^{\frac{1-4\xi_{\mathrm{S}}}{\xi_{\mathrm{S}}}}
+\frac{4\,D\,(D-1)}{l^2}(\xi_{\mathrm{S}}-\xi_D)
(\xi_{\mathrm{S}}-\xi_{D+1})\right),
&\lambda=\vec{k}^2.
\label{stealthpot}
\end{align}
\end{subequations}
For $k_u=0=\vec{k}$, the solution is a holographic massive free
stealth with mass
\begin{equation}\label{eq:massSS}
m_{\mathrm{S}}^2=\frac{2(z-1)(2z-1)\xi_{\mathrm{S}}}{l^2},
\end{equation}
however, this is far from being the more general situation. If
$\vec{k}\ne0$, the solution is in fact stationary since time
dependence can be eliminated via a Galilean boost which is one
of the symmetries underlying the  Schr\"odinger metric
(\ref{eq:Schrometric}), see Ref.~\cite{AyonBeato:2011qw}. This
explains why the coupling constant $\lambda$ is independent of
the constant $k_u$. Using the invariance under spatial
rotations the vector $\vec{k}$ can be aligned along one of the
spatial directions $x^i$. In turn, the constant $\sigma_0$ can
be eliminated exploiting the translation invariance (along
space or time). In summary, there are no independent
integration constants if $\vec{k}\ne0$. For $\vec{k}=0$ and
$k_u\ne0$, the solution describes a time-dependent massive free
stealth overflying the Schr\"odinger background with the same
mass (\ref{eq:massSS}). Notice that the constant $k_u$ can be
tuned to any predetermined value using the dynamical scaling
symmetry of the Schr\"odinger spacetime, and this case is also
free of any independent integration constants.

The above configuration becomes enhanced for the dynamical
scaling $z=1/2$ due to the emergence of conformal symmetry.
Indeed, for $z=1/2$ the Schr\"odinger spacetimes have the
special property that their Weyl tensor vanishes, i.e.\ they
turn into conformally flat spacetimes for this value. This is
manifest by rewriting the metric (\ref{eq:Schrometric}) for
$z=1/2$ as
\begin{align}
ds_{\mathrm{S}}^2&=\frac{l^2}{y^2}\left(
-\frac{y}{l}du^2-2dudv
+dy^2+d\vec{x}^2\right)\nonumber\\
&=\frac{l^2}{y^2}\left\{
-2\,du\,d\left(v+\frac{u^3}{24l^2}+\frac{uy}{2l}\right)
+\left[d\left(y+\frac{u^2}{4l}\right)\right]^2+d\vec{x}^2\right\}
\equiv\Omega^2\eta_{\mu\nu}d\bar{x}^{\mu}d\bar{x}^{\nu},\label{eq:SCF}
\end{align}
where $\{\bar{x}^{\mu}\}$ are the standard cartesian
coordinates of flat spacetime, defined here from the above
light-cone representation. The enhancement occurs because the
Schr\"odinger nonminimal coupling (\ref{xiSchrod}) becomes the
conformal one (\ref{eq:xi_D}) for $z=1/2$, that is
$\xi_{\mathrm{S}}=\xi_D$, which in turn causes that the
conformally invariant potential also emerges from
self-interaction (\ref{stealthpot}). In other words, the
stealth action (\ref{eq:action}) becomes conformally invariant
and having a concrete example of stealth configuration implies
that its whole conformal class also allows a stealth
interpretation. Hence, due to the conformally flat character of
the Schr\"odinger spacetimes for $z=1/2$, their stealth
configurations in this case can be obtained from a conformal
transformation of the stealths defined on flat spacetime
\cite{AyonBeato:2005tu} (a similar mechanism works for the
conformal stealths allowed for any standard cosmology
\cite{Ayon-Beato:2013bsa}). More concretely, the conformal
transformation between the $z=1/2$ Schr\"odinger stealths
$\Phi_{\mathrm{S}}$ and the flat configurations
$\Phi_{\mathrm{F}}$ \cite{AyonBeato:2005tu}, inferred from the
conformal relation (\ref{eq:SCF}), is
\begin{subequations}\label{eq:CSS}
\begin{align}
\Phi_{\mathrm{S}}(x^{\mu}) &= \Omega^{-(D-2)/2}\,
\Phi_{\mathrm{F}}(\bar{x}^{\mu})
 =\left[\Omega\left(\frac{\alpha}2\eta_{\mu\nu}\bar{x}^{\mu}\bar{x}^{\mu}
  +k_{\mu}\bar{x}^{\mu}+\sigma_{0}\right)\right]^{-(D-2)/2}
\nonumber\\
&= \Biggl( \frac{l}{y} \Biggl\{
    \frac{\alpha}2\left[-2\,u\left(v+\frac{uy}{2l}+\frac{u^3}{24l^2}\right)
  + \left(y+\frac{u^2}{4l}\right)^2+\vec{x}^2\right]
  + k_uu
  + k_v\left(v+\frac{uy}{2l}+\frac{u^3}{24l^2}\right)
  \nonumber\\
&\qquad\quad
  + k_y\left(y+\frac{u^2}{4l}\right)
  + \vec{k}\cdot\vec{x}
  + \sigma_0           \Biggr\}
   \Biggr)^{-(D-2)/2}.
\end{align}
Remarkably, explicitly solving the stealth constraints
(\ref{eq:SEqs}) for $z=1/2$ Schr\"odinger spacetimes, in a
lengthy process, gives exactly the same result. The relation
between the conformal coupling constant and the resulting
integration constants is the same than in flat spacetime
\cite{AyonBeato:2005tu}
\begin{equation}\label{eq:CpotS}
U(\Phi) =\frac{(D-2)^2}{8}\lambda\Phi^{\frac{2D}{D-2}}, \qquad
\lambda=-2 k_u k_v+k_y^2+\vec{k}^2-2\alpha\sigma_0.
\end{equation}
\end{subequations}
In flat spacetime and for $\alpha\ne0$ one can use the translation
invariance to put the constants $k_\mu$ to zero all ;
consequently, the conformal stealth depends on a single
independent integration constant (the other one is determined
by the coupling constant), related to the existence of the
conformal symmetry. In the Schr\"odinger case the translation
invariance along the coordinate $y$ is lost and we can only choose
$k_u$, $k_v$ and $\vec{k}$ to be zero. However, the constant
$k_y$ is not arbitrary since it can be fixed by an anisotropic
scaling. Consequently, the conformal stealth of $z=1/2$
Schr\"odinger spacetime has again a single independent
integration constant as its conformal cousin of flat spacetime.
This is not always the case for conformally flat
configurations, as it is evidenced by the cosmological
configurations \cite{Ayon-Beato:2013bsa}, where usually the
conformal factor breaks translation invariance increasing the
number of independent integration constants. The difference here
is that this breaking is compensated by the existence of the
scaling symmetry. Something similar occurs for the isotropic
case $z=1$ which is also conformally flat defining AdS space
\cite{Ayon-Beato:SAdS}. For $\alpha=0$ and $\vec{k}\ne0$ then
$k_u$ and $\sigma_0$ can be put to be zero by a Galilean boost and
a translation, respectively. The constant $k_v$ becomes fixed
by a scaling, and a rotation is responsible of aligning the
vector $\vec{k}$ along a particular spatial direction.
Consequently, the solution again depends on a single
independent integration constant. For $\alpha=0=\vec{k}$, if
$k_u\ne0$ or $k_v\ne0$ then one of them is fixed by a scaling
and $\sigma_0$ can be chosen to be zero by a translation. The
solution would depend on a single independent integration
constant if both $k_u$ and $k_v$ are nonvanishing, and also if
both are vanishing (since $\sigma_0$ would remain free). If
only one of them vanishes there is no independent integration
constant at all. This is also the case for $\alpha=0=k_\mu$,
which coincides just with the $z=1/2$ holographic massless free
stealth solution (\ref{eq:gSS}).

We mentioned at the beginning that the dynamical exponent $z=0$
is exceptional in the sense that the nonminimal coupling is not
necessarily restricted in this case. That is, for $z=0$ the
solution (\ref{eq:gSS}) is still valid and the Schr\"odinger
nonminimal coupling becomes $\xi_{\mathrm{S}}=\xi_{D-1}$,
however, the following additional branch occurs if the
nonminimal coupling is allowed to take any other different
value
\begin{subequations}\label{eq:z=0SS}
\begin{align}
z &=0,\\
\Phi(x^\mu) &= \left[\frac{y}{l}\left(\frac{k_u}{\omega}\sin(\omega{u})
+\sigma_0\cos(\omega{u})\right)^{-1}\right]^{\frac{2\xi}{1-4\xi}},
&\omega^2 = \frac{(D-2)(\xi-\xi_{D-1})}{\xi{l}^2},\\
U(\Phi) &= \frac{8\,D\,(D-1)\xi(\xi-\xi_D)
(\xi-\xi_{D+1})}{(1-4\xi)^2l^2}\,\Phi^2.
\end{align}
\end{subequations}
In the limit $\xi\rightarrow\xi_{D-1}$ the frequency $\omega$
vanishes and we recover the solution (\ref{eq:gSS}) with
$\vec{k}=0$. Here, the constant $\sigma_0$ can be eliminated by
a time translation and the constant $k_u$ can be fixed by a
scaling (since the retarded time $u$ remains unchanged for
$z=0$); once again the stealth has no independent integration
constants.

Finally, notice that $\xi_{\mathrm{S}}<1/4$ in the solution
(\ref{eq:gSS}) and in the limit $\xi\rightarrow1/4$ there
is no redefinition allowing the solution (\ref{eq:z=0SS}) to
have the well-behaved behavior defined in (\ref{eq:hatsigma});
the value $\xi=1/4$ is excluded from the presented solutions. A
careful study of this case shows that like in the other
paradigmatic example of anisotropic background, i.e.\ the
Lifshitz spacetime, no stealth configuration on the
Schr\"odinger spacetime exists for the nonminimal coupling
$\xi=1/4$ if one consider a generic value of the dynamical
exponent $z$. But, contrary to the Lifshitz example, there is
no exceptional anisotropic value of the exponent allowing
solutions for this coupling.

\subsection{Schr\"odinger inspired hyperscaling violation}

As in the standard hyperscaling violation case
(\ref{eq:HSVmetric}), we would like to exploit the obvious
conformal relation between the Schr\"odinger background
(\ref{eq:Schrometric}) and those exhibiting hyperscaling
violation inspired by the Schr\"odinger line element
(\ref{eq:HVSchro}). Hence, we start by using a conformal
argument to map stealth configurations on both backgrounds.
However, as we have emphasized previously, in order to perform
this task it is vital for the scalar field to be a power of the
conformal factor depending, this time, exclusively on the
coordinate $y$. Since, unlike the Lifshitz case, the
Schr\"odinger stealths are not necessarily holographic, these
configurations are obtained by imposing the conditions
$\alpha=0=k_\mu$ in the first two examples (\ref{eq:gSS}) and
(\ref{eq:CSS}). They describe in general holographic massive
free stealths with mass (\ref{eq:massSS}). In order to use the
conformal mapping, the first step is to show that the
transformations (\ref{eq:map}) between two massless free
nonminimally coupled actions (\ref{eq:actionmap}), also relate
the actions when self-interactions are present. Concretely, the
formula (\ref{eq:actionmap}) can be extended to
\begin{subequations}
\begin{equation}
\frac1{\xi}\left(S_{\xi}^0[\Phi,g_{\mu\nu}]
-\int{d^{D}x}\sqrt{-g}\,U(\Phi)\right)=
\frac1{\bar{\xi}}\left(S_{\bar{\xi}}^0[\bar{\Phi},\bar{g}_{\mu\nu}]
-\int{d^{D}x}\sqrt{-\bar{g}}\,\bar{U}(\bar{\Phi})\right)+\rm{b.t.},
\end{equation}
where the self-interactions must be related by
\begin{equation}\label{eq:self-int_map}
\bar{U}(\bar{\Phi})=\frac{\bar{\xi}}{\xi}
\left({\bar{\Phi}}/{\Phi_0}\right)^{\frac{2D\alpha}{(D-2)\alpha-2}}
U\!\!\left(\Phi_0
\left({\bar{\Phi}}/{\Phi_0}\right)^{\frac{2}{2-(D-2)\alpha}}
\right).
\end{equation}
\end{subequations}
For example, starting with a potential given by a superposition
of power laws of the scalar field as
$U(\Phi)=\sum_i\lambda_i\Phi^{\sigma_i}$, the new potential
results again in a superposition of power laws of the new
scalar field where the new parameters are defined by
\begin{equation}
\bar{\lambda}_i=\lambda_i\frac{\bar{\xi}}{\xi}\Phi_0^{
\frac{\alpha(D-2)\sigma_i-2\alpha D}{\alpha(D-2)-2}}, \qquad
\bar{\sigma}_i=\frac{2(\alpha D-\sigma_i)}{\alpha(D-2)-2}.
\end{equation}
Applying this last version of the map to the holographic
massive free configurations (\ref{eq:gSS}), a straightforward
computation shows that the holographic self-interacting
configuration given by
\begin{subequations}\label{HVSconf}
\begin{align}
\xi_{\mathrm{HS}}&\equiv\frac{1}{4}\frac{(D-2)(\theta-2z-D+3)^2}
{(D-1)(\theta-2z-D+3)^2-(2z-1)(2z+D-3)},\label{eq:xi_HS}\\
\Phi(y)&=\left[\left(l\sqrt{\lambda}\right)^{\frac{D-2}{\theta}}
\frac{y}{l}\right]^{\frac{2z+D-3-\theta}{2}},\\
U(\Phi)&=(z-1)(2z-1)\,\xi_{\mathrm{HS}}\,\lambda\,
\Phi^{\frac{2[D\theta-(D-2)(2z+D-3)]}
{(D-2)(\theta-2z-D+3)}},\frac{}{}\label{eq:HVSh_pot}
\end{align}
\end{subequations}
satisfies the stealth constraints (\ref{eq:SEqs}) on the
backgrounds exhibiting hyperscaling violation \emph{\`a la}
Schr\"odinger (\ref{eq:HVSchro}). The constant appearing in the
transformation (\ref{eq:map}) is naturally defined here in
terms of the integration constant of the starting holographic
configuration (\ref{eq:gSS}) as
$\Phi_0^2=\sigma_0^{-(2z+D-3)}$. This constant can be fixed
arbitrarily on the Schr\"odinger background using a scaling,
however, this is no longer the case in the hyperscaling
violation context since now it defines the coupling constant of
the self-interaction (\ref{eq:HVSh_pot}) via $\Phi_0^2 =
(l\sqrt{\lambda})^{(D-2)(2z+D-3-\theta)/\theta}$. We would like
to stress that the utility of the above approach, beyond its
succinctness and beauty, is that if one explicitly solves the
stealth constraints (\ref{eq:SEqs}) for generic values of the
exponents characterizing these backgrounds the only
configuration valid for all the cases is precisely the above.
As in the standard hyperscaling violation case, this is not the
end of the story since there are also special values of the
exponents for which the behavior of the stealths can be
different from this conformally generated holographic
configuration.

We start the covering of especial cases by pointing out that
there are only two special values of the dynamical exponent $z$
for which the Schr\"odinger hyperscaling violation nonminimal
coupling (\ref{eq:xi_HS}) becomes the conformal one
(\ref{eq:xi_D}), $\xi_{\mathrm{SH}}=\xi_D$. The first, is again
the point $z=1/2$ as in the purely Schr\"odinger analysis, and
the second is for the exponent $z=-(D-3)/2$. For $z=1/2$ the
self-interaction (\ref{eq:HVSh_pot}) vanishes, but
nonholographic contributions emerge in this case and supplement
the self-interaction precisely with the conformal potential; as
a direct consequence, the stealth action (\ref{eq:action}) is
again conformally invariant. Due to the conformal relation of
these hyperscaling violation backgrounds to the Schr\"odinger
one (\ref{eq:HVSchro}), the resulting conformal configuration
is just a conformal transformation of the Schr\"odinger
conformal stealth (\ref{eq:CSS})
\begin{equation}
\Phi_{\mathrm{HS}}=\left(\frac{l}{y}\right)^{\frac\theta2}
\Phi_{\mathrm{S}},
\end{equation}
where the conformal coupling constant is determined as in
Eq.~(\ref{eq:CpotS}). These solutions allow all the subcases
already characterized for the Schr\"odinger example in the
paragraph following Eq.~(\ref{eq:CpotS}). For the exponent
$z=-(D-3)/2$ not only the Schr\"odinger hyperscaling violation
nonminimal coupling (\ref{eq:xi_HS}) becomes the conformal
coupling, but also the self-interaction (\ref{eq:HVSh_pot})
turns out to be the conformal one. Nothing special happens in
this case, since (\ref{HVSconf}) just describes a conformal
holographic stealth. It is curious that this stealth can be
obtained as a conformal transformation of the constant scalar
trivial solution allowed by the Schr\"odinger AdS-wave
\begin{equation}
ds_{\mathrm{S}}^2=\frac{l^2}{y^2}\left[
-\left(\frac{y}{l}\right)^{D-1}du^2-2dudv
+dy^2+d\vec{x}^2\right],
\end{equation}
which happens to be the holographic vacuum solution of Einstein
equations in the presence of a negative cosmological constant
\cite{AyonBeato:2011qw}, i.e.\ in this case
$\Phi_{\mathrm{HS}}=({l}/{y})^{\frac\theta2}
(l\sqrt{\lambda})^{\frac{2-D}{2}}$.

The other special cases appear when the exponents are taken as
$\theta=D-2$ and $z=0$, which reduce the Schr\"odinger
hyperscaling violation backgrounds (\ref{eq:HVSchro}) to the
\emph{pp}-wave
\begin{equation}
ds_{\mathrm{HS}}^2=-\left(\frac{y}{l}\right)^2du^2-2dudv
+dy^2+d\vec{x}^2.
\end{equation}
The following purely time-dependent massless free stealth
\begin{subequations}\label{eq:hsptd}
\begin{align}
z &=0,\frac{}{}\\
\theta &=D-2,\frac{}{}\\
\Phi(u) &= \left[A\sin\left(\sqrt{\frac{1-4\xi}{4\xi}}\frac{u}l\right)
+B\cos\left(\sqrt{\frac{1-4\xi}{4\xi}}\frac{u}l\right)
\right]^{-\frac{2\xi}{1-4\xi}},
\end{align}
\end{subequations}
overflies the previous \emph{pp}-wave for a generic value of
the nonminimal coupling parameter $\xi$. There are two special
nonminimal couplings for which this purely time-dependent
configuration acquires an additional dependence along the
holographic direction $y$. Firstly, for the Schr\"odinger
hyperscaling violation nonminimal coupling (\ref{eq:xi_HS}),
which becomes $\xi=\xi_{\mathrm{HS}}=1/8$ for the exponents
$\theta=D-2$ and $z=0$, the additional contribution gives rise
to a self-interacting stealth
\begin{subequations}
\begin{align}
z &=0,\frac{}{}\\
\theta &=D-2,\frac{}{}\\
\xi &=\frac18,\\
\Phi(x^\mu) &= \frac1{\sqrt{A\sin\left(\frac{u}l\right)
+B\cos\left(\frac{u}l\right)
+\sqrt{\lambda}y}},\\
U(\Phi) &= \frac{\lambda}8\,\Phi^6.
\end{align}
\end{subequations}
The other case allowing an additional dependence along the
holographic direction is for the conformal coupling,
$\xi=\xi_D$, since the hyperscaling violation exponent $\theta$
is not necessarily restricted in this case and the extra
dependence enters via a nontrivial conformal factor
\begin{subequations}
\begin{align}
z &=0,\\
\xi &=\xi_D=\frac{D-2}{4(D-1)},\\
\Phi(x^\mu) &= \left(\frac{l}{y}\right)^{\frac{\theta-D+2}2}
\left[A\sin\left(\frac{u}{l\sqrt{D-2}}\right)
+B\cos\left(\frac{u}{l\sqrt{D-2}}\right)\right]^{-\frac{D-2}{2}}.
\end{align}
\end{subequations}
In all these time-dependent solutions the constant $B$ can be
eliminated by a time translation and these solutions have a
single independent integration constant.

Finally, we comment on the nonminimal coupling value $\xi=1/4$,
which  deserves special attention. First, we notice that
Schr\"odinger hyperscaling violation nonminimal coupling
(\ref{eq:xi_HS}) contains this value, $\xi_{\mathrm{HS}}=1/4$,
for
\begin{equation}
z=\frac{\theta^2-(D-3)[2\theta-(D-2)]}
{2(2\theta-D+2)}.
\end{equation}
Hence, the behavior of the stealths with $\xi=1/4$ for a
generic value of the hyperscaling violation exponent $\theta$
is ruled by the holographic configuration (\ref{HVSconf})
restricting the dynamical exponent $z$ as above. An exception
must be made again for the exponents $\theta=D-2$ and $z=0$
where the holographic behavior is broken according to
\begin{subequations}
\begin{align}
z &=0,\frac{}{}\\
\theta &=D-2,\frac{}{}\\
\xi &=\frac14,\\
\Phi(x^\mu) &= \exp\left(\frac{u^2}{4l^2}+k_uu+\vec{k}\cdot\vec{x}
+\sigma_0\right),\\
U(\Phi) &=\frac12\,m^2\,\Phi^2,
&m^2=\vec{k}^2.
\end{align}
\end{subequations}
The above massive configurations with $\vec{k}\ne0$ have no
independent integration constants since $k_u$ and $\sigma_0$
can be chosen to be zero using translations, and $\vec{k}$ can be
rotated to a particular spatial axis where the related
component just determines the mass. We found intriguing that
contrary to the previous $\xi=1/4$ solutions we exhibited, this
configuration can not be obtained from a solution with generic
values of the nonminimal coupling parameter via the nontrivial
limit described in Subsec.~\ref{subsec:Lz=0}. This is possible
only in the massless case $\vec{k}=0$ which is a nontrivial
limit of the purely time-dependent solution (\ref{eq:hsptd}).
The limit inherited the property that only one constant can be
eliminated by time translation, i.e.\ these stealths overfly
the involved background with a single independent integration
constant.

\section{Conclusions}

Here, we have established the existence of stealth
configurations given by scalar fields nonminimally coupled to
the gravity of some interesting anisotropic backgrounds such as
the Lifshitz and Schr\"odinger metrics as well as their hyperscaling
violation extensions. These backgrounds have been proposed
recently to extend the ideas underlying the AdS/CFT
correspondence to non-relativistic field theory.

We have first shown that Lifshitz stealths with nontrivial
dynamical exponent $z$, because of the anisotropy, are possible
only as massless free scalar fields depending exclusively on
the holographic coordinate. Moreover, the nonminimal coupling
is not arbitrary but is parameterized in terms of the dynamical
exponent $z$ of the Lifshitz spacetime for any dimension. All
these features make the Lifshitz stealth configurations
different from those existing in the $z=1$ isotropic (A)dS case
\cite{Ayon-Beato:SAdS} and in the Minkowski spacetime
\cite{AyonBeato:2005tu}, since those latter are generally
self-gravitating, depend on all the coordinates and exist for
all values of the nonminimal coupling parameter. An exception
must be made for the vanishing exponent $z=0$ where Lifshitz
spacetime becomes conformally flat and additional time
dependencies appear. Interestingly, within this context, we
provide for the first time an approach to obtain stealth
solutions with a nonminimal coupling $\xi=1/4$ from the
solutions with a generic value of the coupling which usually
exclude this case. The procedure seems to give rise to the more
general solutions in most of the cases.

We have taken advantage of the fact that the hyperscaling
violation metric is conformally related to the Lifshitz one as
well as of the holographic character of the Lifshitz stealth
configuration to obtain a stealth configuration defined on the
hyperscaling violation background. This has been done by
mapping the Lifshitz stealth solution using a very simple
conformal argument relating the actions of two different
nonminimally coupled massless free theories. We have shown that
the resulting configuration is just a particular case of the
stealth solutions defined on the hyperscaling violation metric
since we have also derived the most general stealth
configurations which exhibit a non-holographic character for
special values of the exponents. In the last section, we have
also considered the case of Schr\"odinger stealth
configurations and shown the character not necessarily
holographic of the stealth solutions. This difference is
essentially due to the presence of a null direction in the
Schr\"odinger metric. These solutions can be made holographic
by fixing appropriately the constants and, extending the
conformal argument of the previous case to self-interacting
actions, we have mapped these Schr\"odinger holographic stealth
into stealth solutions defined on the Schr\"odinger background
with hyperscaling violation. We have scanned also the special
values of the exponents for which the stealth behaviors depart
from being holographic.

As a natural but highly nontrivial work, it would be
interesting to characterize geometrically all the static
spacetimes that may support stealth configurations given by a
nonminimal scalar field. The spacetime geometries considered in
this paper are all of zero mass. A possible extension of the
present work is to look for stealth configurations defined on
black holes having as asymptotic the studied anisotropic
geometries. These spacetimes play an important role since they
are the gravitational duals at finite temperature regime of
nonrelativistic gauge/gravity duality. Examples of stealths on
black holes are known on the BTZ one \cite{AyonBeato:2004ig},
on the $z=1$ black hole of New Massive Gravity
\cite{Hassaine:2013cma}, and for a more general scalar tensor
theory in the case of the Schwarzschild metric or for Lifshitz
black holes, see \cite{Babichev:2013cya,Bravo-Gaete:2013dca}.
There are even examples of composite stealths on the BTZ black
holes \cite{Cardenas:2014kaa}.

Another interesting task would consist in finding some
applications concerning these stealth configurations in the
set-up of the non-relativistic version of the AdS/CFT
correspondence. A starting point is studying the perturbations
of all the configurations we present here. It is possible that
most of them are not stealth themselves due to the rigidity of
stealth solutions, i.e.\ they do not present integration
constants or have the smallest possible number of them, causing
the perturbations to evolve necessarily departing from the
stealth regime. The coupling of stealth perturbations to the
ones of the gravitational duals unavoidably would change the
behavior of the last ones bringing nontrivial implications for
holographic predictions. We hope to explore this interesting
issue in the near future.

\begin{acknowledgments}
The authors thank M. Bravo-Gaete and useful discussions with C.~Terrero-Escalante.
EAB is partially supported by grants 175993 and 178346 from
CONACyT, grants 1121031, 1130423 and 1141073 from FONDECYT and
``Programa Atracci\'{o}n de Capital Humano Avanzado del
Extranjero, MEC'' from CONICYT. MH is partially supported by
grant 1130423 from FONDECYT. This project was partially funded
by Proyectos CONICYT- Research Council UK - RCUK -DPI20140053.
\end{acknowledgments}



\begin{thebibliography}{99}

\bibitem{Banados:1992wn}
  M.~Banados, C.~Teitelboim and J.~Zanelli,
  Phys.\ Rev.\ Lett.\  {\bf 69}, 1849 (1992).

\bibitem{AyonBeato:2004ig}
  E.~Ayon-Beato, C.~Martinez and J.~Zanelli,
  Gen.\ Rel.\ Grav.\  {\bf 38}, 145 (2006).

\bibitem{AyonBeato:2005tu}
  E.~Ayon-Beato, C.~Martinez, R.~Troncoso and J.~Zanelli,
  Phys.\ Rev.\ D {\bf 71}, 104037 (2005).

\bibitem{Ayon-Beato:SAdS} E.~Ay\'{o}n--Beato,
    C.~Mart\'{\i}nez, R.~Troncoso, and J.~Zanelli,
    ``Stealths overflying (A)dS,'' in preparation.

\bibitem{Ayon-Beato:2013bsa}
  E.~Ay\'{o}n-Beato, A.~A.~Garc\'{\i}a, P.~I.~Ram\'{\i}rez-Baca and C.~A.~Terrero-Escalante,
  Phys.\ Rev.\ D {\bf 88}, 063523 (2013).

\bibitem{Maldacena:1997re}
  J.~M.~Maldacena,
  Adv.\ Theor.\ Math.\ Phys.\  {\bf 2}, 231 (1998).

\bibitem{Kachru:2008yh}
  S.~Kachru, X.~Liu and M.~Mulligan,
  Phys.\ Rev.\ D {\bf 78}, 106005 (2008).

\bibitem{Charmousis:2010zz}
  C.~Charmousis, B.~Gouteraux, B.~S.~Kim, E.~Kiritsis and R.~Meyer,
  JHEP {\bf 1011}, 151 (2010).

\bibitem{Son:2008ye}
  D.~T.~Son,
  Phys.\ Rev.\ D {\bf 78}, 046003 (2008).

\bibitem{Balasubramanian:2008dm}
  K.~Balasubramanian and J.~McGreevy,
  Phys.\ Rev.\ Lett.\  {\bf 101}, 061601 (2008).

\bibitem{Taylor:2008tg}
  M.~Taylor,
  arXiv:0812.0530 [hep-th].

\bibitem{AyonBeato:2009nh}
  E.~Ayon-Beato, A.~Garbarz, G.~Giribet and M.~Hassaine,
  Phys.\ Rev.\ D {\bf 80}, 104029 (2009).

\bibitem{AyonBeato:2010tm}
  E.~Ayon-Beato, A.~Garbarz, G.~Giribet and M.~Hassaine,
  JHEP {\bf 1004}, 030 (2010).

\bibitem{Alvarez:2014pra}
  A.~Alvarez, E.~Ay\'{o}n-Beato, H.~A.~Gonz\'{a}lez and M.~Hassa\"{\i}ne,
  JHEP {\bf 1406}, 041 (2014).

\bibitem{HyperBH}
  M.~Cadoni and M.~Serra, JHEP {\bf 1211}, 136 (2012);
  M.~Alishahiha, E.~O Colgain and H.~Yavartanoo, JHEP {\bf 1211}, 137 (2012);
  P.~Bueno, W.~Chemissany, P.~Meessen, T.~Ortin and C.~S.~Shahbazi,
  JHEP {\bf 1301}, 189 (2013);
  M.~Hassaine,
  Phys.\ Rev.\ D {\bf 91}, 084054 (2015);
  M.~Bravo-Gaete, S.~Gomez and M.~Hassaine,
  arXiv:1505.00702 [hep-th].

\bibitem{Dong:2012se}
  X.~Dong, S.~Harrison, S.~Kachru, G.~Torroba and H.~Wang,
  JHEP {\bf 1206} (2012) 041.

\bibitem{Duval:2008jg}
  C.~Duval, M.~Hassaine and P.~A.~Horvathy,
  Annals Phys.\  {\bf 324}, 1158 (2009).

\bibitem{AyonBeato:2006jf}
  E.~Ayon-Beato and M.~Hassaine,
  Phys.\ Rev.\ D {\bf 75}, 064025 (2007).

\bibitem{AyonBeato:2005qq}
  E.~Ayon-Beato and M.~Hassaine,
  Phys.\ Rev.\ D {\bf 73}, 104001 (2006)
  [hep-th/0512074].

\bibitem{AyonBeato:2011qw}
  E.~Ayon-Beato, G.~Giribet and M.~Hassaine,
  Phys.\ Rev.\ D {\bf 83}, 104033 (2011)
  [arXiv:1103.0742 [hep-th]].

\bibitem{Kim:2012nb}
  B.~S.~Kim,
  JHEP {\bf 1206}, 116 (2012).

\bibitem{Hassaine:2013cma}
  M.~Hassaine,
  Phys.\ Rev.\ D {\bf 89}, 044009 (2014).

\bibitem{Babichev:2013cya}
  E.~Babichev and C.~Charmousis,
  JHEP {\bf 1408}, 106 (2014).

\bibitem{Bravo-Gaete:2013dca}
  M.~Bravo-Gaete and M.~Hassaine,
  Phys.\ Rev.\ D {\bf 89}, 104028 (2014).

\bibitem{Cardenas:2014kaa}
  M.~Cardenas, O.~Fuentealba and C.~Mart\'{\i}nez,
  Phys.\ Rev.\ D {\bf 90}, 124072 (2014).

\end{thebibliography}
\end{document}